\documentclass[%
 square,
 amsfonts,
 reprint,
 amsmath,amssymb,
 aps, superscriptaddress,
]{revtex4-2}

\usepackage{graphicx}%
\usepackage{dcolumn}%
\usepackage{bm}%
\usepackage[utf8]{inputenc}
\usepackage[hmargin=1.0cm]{geometry}
\usepackage{listings}	%
\usepackage{amsmath}
\usepackage{amssymb}
\usepackage[caption=false]{subfig}
\usepackage{amsthm}	%
\usepackage{todonotes}	%
\usepackage{marginnote}	%
\usepackage{siunitx}	%
\usepackage{slashed}	%
\usepackage{mathrsfs}	%
\usepackage[nolist]{acronym}
\usepackage[colorlinks=true, allcolors = blue]{hyperref}	%
\usepackage[normalem]{ulem}
\renewcommand{\v}[1]{{\ensuremath{\boldsymbol{\mathbf{#1}}}}} %
\newcommand{\gv}[1]{\ensuremath{\mbox{\boldmath$ #1 $}}} 
\newcommand{\abs}[1]{\left| #1 \right|} %

\renewcommand{\Re}{\operatorname{Re}}
\renewcommand{\Im}{\operatorname{Im}}

\newcommand{\rom}[1]{%
  \textup{\uppercase\expandafter{\romannumeral#1}}  %
}
\newcommand*\chem[1]{\ensuremath{\mathrm{#1}}}

\DeclareMathOperator{\sgn}{sgn}

\let\OLDthebibliography\thebibliography
\renewcommand\thebibliography[1]{
  \OLDthebibliography{#1}
  \setlength{\parskip}{0pt}
\setlength{\itemsep}{0pt plus 0.3ex}
}

\begin{document}
\title{Thermal fluctuations and vortex lattice structures in chiral $p$-wave superconductors: robustness of double-quanta vortices.}
\author{Fredrik Nicolai Krohg}
\affiliation{\footnotesize Department of Physics, Norwegian University of Science and Technology, NO-7491, Trondheim, Norway}
\affiliation{\footnotesize Center for Quantum Spintronics, Department of Physics, Norwegian University of Science and Technology, NO-7491, Trondheim, Norway}
\author{Egor Babaev}
\affiliation{Department of Physics, KTH-Royal Institute of Technology, Stockholm SE-10691, Sweden}
\author{Julien Garaud}%
\affiliation{\footnotesize Institut Denis Poisson CNRS-UMR 7013,  
            Universit\'e de Tours, 37200 France}
\author{H\aa vard Homleid Haugen}
\author{Asle Sudb\o}
\affiliation{\footnotesize Department of Physics, Norwegian University of Science and Technology, NO-7491, Trondheim, Norway}
\affiliation{\footnotesize Center for Quantum Spintronics, Department of Physics, Norwegian University of Science and Technology, NO-7491, Trondheim, Norway}
\date{\today}%

\begin{abstract}
We use large-scale Monte-Carlo simulations to study {thermal fluctuations in chiral $p$-wave superconductors  in an applied magnetic field in three dimensions.
We consider the thermal stability of previously predicted unusual double-quanta flux-line lattice ground states in such superconductors.
In previous works it was shown that, neglecting thermal fluctuations, a chiral $p$-wave superconductor forms an hexagonal  lattice of doubly-quantized vortices, except extremely close to the vicinity of $H_{c2}$ where double-quanta vortices split apart. 
We find dissociation of double-quanta vortices driven by thermal fluctuations. However, our calculations also
show that the previous predictions of hexagonal doubly-quantized vortices, where thermal fluctuations were ignored, are very robust
in the considered model.}
\end{abstract}

\maketitle
\section{Introduction}\label{sec:intro}

Higher angular momentum odd-parity chiral superfluid and superconducting states are highly non-trivial pairing symmetries that result in novel topological as well as thermodynamic properties. 
Examples are chiral $p$-wave and chiral $f$-wave states.  
A prominent example of a condensed matter system where such a phase is firmly established, is within the very rich phase diagram of superfluid  \chem{^3He}, where the so-called $A$-phase is a chiral $p$-wave superfluid.
This unconventional superfluid phase was first discovered in seminal works of Osheroff {\it et al}  \cite{Osheroff72,OsheroffPressure72,Leggett74,Leggett75}.   
{It is the interplay between spin- and orbital degrees of freedom, with the multi-component nature of the matter field of the superfluid or superconducting states, that makes the physics of such condensates much richer than the corresponding physics in simple superfluids like \chem{^4He}}
\footnote{Note that we deliberately avoid the use of the term  order parameter when discussing the matter field. 
    The reason is, while the identification of the matter field as a local order parameter would be perfectly fine in a $3D$ superfluid, this is not so for a superconductor. 
    In a superconductor, with a charged condensate coupling minimally to a gauge-field, there exists no local order parameter in any dimension, at any temperature \cite{Elitzur}. 
    Throughout our paper, we simply refer to what is normally called the superconducting order parameter, as a {\it field}. 
{Under certain circumstances the
Higgs mass of the gauge-field acquired upon entering the superconducting state may be interpreted as an order parameter.}
  }. 
The $A$-phase of \chem{^3He} has been used to explain exotic phenomena such as a non-vanishing orbital angular momentum in thermal equilibrium
and unconventional dissipation behaviour due to core-less vortex textures \cite{Mermin76,Anderson77,Mermin79}.

On the other hand, chiral $p$-wave pairing in solid state systems, i.e. superconductors, has remained less well-established. One candidate superconductor with such chiral pairing  that has  been intensely investigated since its discovery, 
is the superconducting phase of \chem{Sr_2RuO_4} \cite{Maeno94}. The crystallographic structure of this compound is a perovskite, similar to the  high $T_c$ cuprates. The normal metallic phase  features transport properties consistent with a $2D$ strongly correlated Fermi liquid phase \cite{Mackenzie1996}, and superconductivity arises out of this normal state at $T \approx 1.5$K. Contrary to the high-$T_c$ cuprates however, \chem{Sr_2RuO_4} is a weak-coupling superconductor. For an early review of the basics physics and superconductivity of \chem{Sr_2RuO_4}, see \cite{Physics_Today_Maeno_2001}. 

{Conventional   pairing  is excluded in \chem{Sr_2RuO_4}
by the many unusual experimental properties of \chem{Sr_2RuO_4}.
Early works revealed a number of unusual features and gave indication of chiral $p$-wave superconductivity.
The early experimental results included the indication of suppression of superconductivity by
non-magnetic impurities \cite{Mackenzie17,Mackenzie98,Mackenzie03}. A conventional superconductor is expected to have a $T_c$ independent of
addition of small fractions of such impurities but rather depend only on the number of magnetic impurities.} 
Early NMR Knight shift experiments showed a temperature-independent Knight-shift and thus a 
residual spin-susceptibility as $T\rightarrow0$, which is a hall-mark of spin-triplet pairing \cite{Ishida98,Ishida15}.
Instead of being isotropic, the gap in \chem{Sr_2RuO_4} is indicated to contain line-nodes or near nodes by both 
the temperature dependence of the specific heat and thermal conductivity 
as well as scanning tunnelling microscopy measurements of the density of states. Other early works on the anisotropy of the thermal conductivity
also were interpreted in favor of chiral $p$-wave pairing state \cite{Izawa01,Nishizaki98,Firmo13}.
Evidence for unconventional  pairing in \chem{Sr_2RuO_4} is provided by the combination of evidence for spontaneous breaking of time-reversal
symmetry and spin-triplet pairing.
Muon spin-relaxation experiments  find spontaneous magnetization in the superconducting state. 
Kerr effect experiments find a temperature dependent Kerr twisting angle \cite{Luke98,Xia06,grinenko2020split} which, significantly, depends on the sign of the magnetic field.

 One of the main predictions of theories  of superconductors with chiral $p$-wave symmetry, is the existence of domains of different chiralities of the superconducting order parameter, and as a result of this, the existence of chiral edge currents between domains of different chirality.
These chiral edge currents should produce magnetic signatures observable by scanning Hall probe microscopy. 
No experimental proof of such chiral edge currents exist, in spite of  several attempts to detect them \cite{Curran14}.
Another issue is that recent \chem{^{17}O} Knight shift results have seen a substantial reduction of spin-susceptibility at low temperatures, 
which led to recently strengthened arguments 
against the hypothesis of spin-triplet pairing \cite{Pustogow19}.
However, the evidence for spontaneous 
symmetry breaking \cite{Xia06,grinenko2020split} , ultrasound \cite{benhabib2020ultrasound},thermodynamics \cite{ghosh2020thermodynamic}
and unconventional vortex physics \cite{ray2014muon}
strongly indicates a multicomponent order parameter. 
 Recent works  suggested the
 possibility of chiral $d$-wave, $s+id$
 and $s+ig$  order parameters for  the superconducting state of \chem{Sr_2RuO_4} \cite{Kivelson20,ghosh2020thermodynamic,romer2019knight}.
 The intense experimental pursuit and controversies associated with
chiral $p$-wave pairing motivates the current work focused on
magnetic response of such systems. Moreover, the model we consider is consistent with a certain chiral $d$-wave order parameter \cite{Kivelson20}, that is presently discussed in connection with \chem{Sr_2RuO_4}.

Furthermore, \chem{UPt_3} is a heavy fermion topological  type-II superconductor with an unconventional superconducting state believed to be a chiral $f$-wave pairing state with \chem{E_{2u}} irreducible representation. At a phenomenological level, it can be
described by a Ginzburg-Landau (GL) theory of a two-component complex matter field with the components related by a time-reversal transformation and oppositely directed internal orbital angular momentum \cite{Sauls90}. The experimental evidence for such a two-component description of the superconducting state of \chem{UPt_3} was recently strengthened when its
time-reversal symmetry breaking character was demonstrated by showing that the energy of the vortex lattice state depends on the relative direction
of the external magnetic field \cite{Sauls20}. The theoretical description we will use is thus relevant to this system.

Early numerical work showed that such a two-component GL theory for \chem{UPt_3} admits anisotropic vortices with
non-trivial core structures and a hexagonal vortex lattice consisting of doubly-quantized vortices at field strengths $H < 0.3 H_{c2}$ in the ground state \cite{SaulsTokuyasu90}.
At higher field strengths $H > 0.3 H_{c2}$, the  doubly-quantized vortices were found to dissociate, into singly-quantized vortices. However, the lattice symmetry of the resulting singly-quantized aggregate vortex state was not determined.

The GL-theory used in this paper, which posits a chiral symmetry of the superconducting state,
is based on the (two-dimensional) $\Gamma^-_{5u}$ representation of the $D_{4h}$ symmetry group \cite{SigristUeda91}. Lowest Landau-level calculations
 based on this GL-theory have predicted a square lattice of vortices when the external magnetic field is applied
parallel to the $c$ axis for high field strengths close to upper critical $H_{c2}$ \cite{AgterbergVortex98}. For fields parallel to the $c$ axis close to the lower critical field $H_{c1}$, an extended London theory predicted a singly-quantized rectangular vortex lattices continuously deforming to singly-quantized square vortex latices as the magnetic field strength was increased \cite{Heeb99}. (Below, we will define precisely what is meant by singly-quantized and doubly-quantized vortices). 
Numerical energy minimization of the free energy has shown that isolated doubly-quantized vortices are {generically} stable and actually are
energetically favorable compared to two isolated single-quanta vortices \cite{Garaud15,garaud2012skyrmionic}.
{In a part of  parameter space, this is corroborated by calculations of isolated topological defects based on Eilenberger's equation where
a $\Gamma^-_{5u}$ symmetry was assumed \cite{Sauls09}. }
{This led to the expectation that  double-quanta vortices form hexagonal lattices, while the
single-quanta vortices form square lattices based on the symmetry of the current distribution of the isolated vortices. }

The numerical studies of isolated vortices were extended to a finite ensemble of vortices in \cite{AsleGaraud16}, where a finite-element
method was used to minimize the GL free energy when increasing the external magnetic field strength.
{These computations found a robust hexagonal lattice of doubly-quantized vortices at field-strengths up to a very close vicinity of $H_{c2}$ when the field was parallel to the $c$-axis. This is
inconsistent with the vortex phase diagram of 
\chem{Sr_2RuO_4} \cite{ray2014muon}.
To examine the vortex structure at fields close to $H_{c2}$, a temperature dependence was inserted into the quadratic coefficient of the free
energy which allowed the system to be moved horizontally in the $T-H$ phase space. 
Extremely close to $H_{c2}$, the double-quanta vortices
were seen to dissociate into   single-quanta vortices that arranged themselves in a square lattice through a mixed phase were both single
and double quanta vortices were present. 
This type of behavior was, on the one hand, quite robust, but
on the other hand has never been observed
in the materials that are candidates for chiral superconductivity.

The manner in which thermal effects were included in \cite{AsleGaraud16} was at a mean field level, i.e.
entropic effects were not fully accounted for.
This then leaves open the question of whether these unusual vortex states and the field-induced transitions between them, are actually stable when  thermal fluctuations are included.
In particular a weak binding energy as well as different entropic contributions of different lattices can alter
the conclusion of the dominant character of two-quanta vortex lattice.} 

In other words, we will investigate in this paper whether the predicted  field-regime of a square singly-quantized vortex lattice with a transition to a doubly-quantized hexagonal vortex lattice \cite{AsleGaraud16} is dramatically over- or underestimated by not fully accounting for entropic effects. 
This is particularly important in this system because 
the mean-field-based Ginzburg-Landau model in an external field, gives two different vortex-lattice states that are  close in free energy. Thus, an assessment of whether the conclusion is robust 
against more accurate estimates of the entropy of the system, is required. Specifically, we attempt to answer if the 
double-quanta vortex lattice survives inclusion of thermal fluctuations since there
is more entropy in a
single-quanta vortex lattice, opening the possibility that it may be entropically stabilized at lower temperatures than the mean-field calculation would predict. Our
approach is related in spirit to that of Ref.~\cite{PhysRevB.82.214512}, where
a decay of single-quanta vortex lattice into a half-quanta lattice was considered
at elevated temperatures.

The purpose of the present paper is therefore to consider the stability of doubly-quantized hexagonal vortex lattices and singly-quantized square vortex lattices when all thermal fluctuation effects are included in gauge-fields and phases of the complex matter fields. 
{ In strongly type-II one-component superconductors,
a good approximation is to neglect amplitude fluctuations\cite{Dasgupta-Halperin,Peskin1978,Kleinert1989gauge,Hove-Mo-Sudbo,Smiseth-Smorgrav-Sudbo,Fossheim-Sudbo-Book}. In chiral superconductors, the situation is more subtle because of a number of massive normal modes that are linear combinations of phase-modes,  magnetic modes,  and amplitude modes \cite{speight2019chiral}.
Then the  London-like approximation
amounts to dropping the most massive modes and neglecting some of the mixing.}

We present the results of extensive Monte-Carlo simulations of a chiral $p$-wave GL-theory with an external field parallel to the $c$-axis. 
This paper is organized as follows. In Section~\ref{sec:GLmodel}, we present in detail the model we will consider, along with a discussion of its parameterization.
We then discuss a subtle point on the discretization of this model on a numerical grid, and the choice of basis for the two-component matter field.
In Section \ref{sec:mc}, we present details of the Monte-Carlo simulations along with definitions of the observables we will use.
In Section \ref{sec:latticeStructures}, we present results of our detailed Monte-Carlo simulations at a filling fraction $f$ of field-induced vortices of $f=1/64$ at various temperatures, starting from high temperatures and proceeding to lower temperatures.
We find two types of stable vortex lattices and an interesting transition region where the vortex lattices thermally reconstruct.
Conclusions are presented in Section \ref{sec:Summary}.
Some mathematical details are relegated to appendices. 

\section{Ginzburg-Landau Model} \label{sec:GLmodel}
\subsection{Dimensionless units and reduction of parameters}

We consider the clean limit of the Ginzburg-Landau energy density
of the two dimensional $\Gamma_{5u}$ irreducible representation of the tetragonal $D_{4h}$
symmetry group which in the chiral basis using dimensionless variables and units reads \cite{AsleGaraud16,Heeb99}
\begin{subequations}
    \label{eq:GLmodel:chiral_clean_limit}
    \begin{align}
        \mathcal{F} =& g^{-2}|\nabla\times\v{A}|^2 + |D_x\v{\eta}|^2 + |D_y\v{\eta}|^2 + 2\tilde{\kappa}_5|D_z\v{\eta}|^2
        \label{eq:GLmodel:chiral_clean_limit:kinetic}\\
        & + (1+\nu)\Re[(D_x\eta_+)^\ast D_x\eta_- - (D_y\eta_+)^\ast D_y\eta_-]
        \label{eq:GLmodel:chiral_clean_limit:anisotropy}\\
        & - (1-\nu)\Im[(D_x\eta_+)^\ast D_y\eta_- + (D_y\eta_+)^\ast D_x\eta_-]
        \label{eq:GLmodel:chiral_clean_limit:mgt}\\
        & + 2|\eta_+\eta_-|^2 + \nu\Re(\eta_+^{\ast 2}\eta_-^2) + \sum_{h=\pm}(-|\eta_h|^2 + \frac{1}{2}|\eta_h|^4).\hspace{-.5em}
        \label{eq:GLmodel:chiral_clean_limit:potential}
    \end{align}
\end{subequations}
The two dimensions of the representation are spanned by the complex fields $\eta_\pm$.
The covariant derivative $D_a = \nabla_a - iA_a$, and $\nu$ and $g$ are dimensionless material parameters with the restriction that $\abs{\nu}<1$.
Deriving the effective Ginzburg-Landau energy from a microscopic model \cite{Krohg18}
, it is seen that $\nu = (\langle v_x^4\rangle - 3\langle v_x^2v_y^2\rangle)/(\langle v_x^4\rangle + \langle v_x^2v_y^2\rangle)$, where $v_a$ is the $a$-component of the Fermi velocity and the brackets: $\langle\cdot\rangle$, indicate an average over the Fermi surface. 
$\nu$ thus parameterizes the anisotropy of the Fermi surface in that $\nu=0$ for a cylindrical surface, while $\nu\neq0$ for a Fermi surface distorted by four-fold anisotropy.

\begin{figure}[h]
    \centering
    \includegraphics{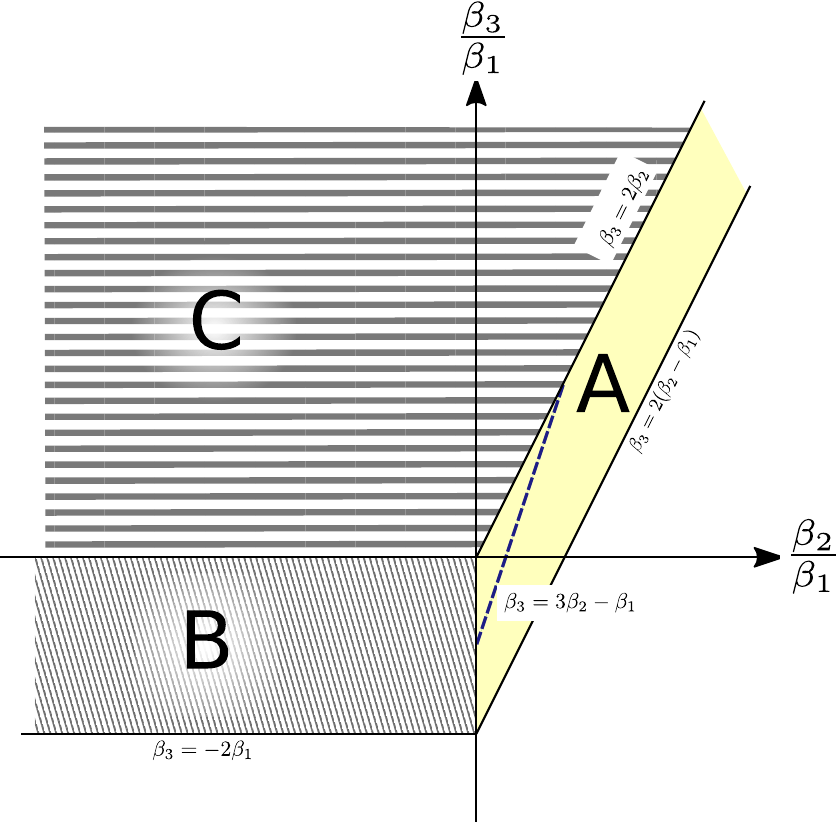}
    \caption{Mean field phase diagram spanned by the fourth order material constants $\beta_i$. 
The A, B and C regions correspond to the mean field solutions given in Table~\ref{tab:GLmodel:MF_phases}. 
The white region below the phases give unbounded mean field energy and is unphysical. The blue line $\beta_3 = 3\beta_2-\beta_1$ gives the values of parameter space spanned by the parameter choices used in the reduced free energy density. 
This line can be parametrized in terms of the single dimensionless parameter $\nu$ for $-1<\nu<1$. }
    \label{fig:GLmodel:MF_phases}
\end{figure}

The model in Eq.~\eqref{eq:GLmodel:chiral_clean_limit} is a restricted version of the full $\Gamma_{5u}$ free energy which in SI units can be written
\cite{SigristUeda91,Agterberg98,Heeb99}
\begin{equation}
    \begin{split}
        \mathcal{F} = &-\alpha\abs{\v{\eta}}^2 + \frac{\beta_1}{2}\abs{\v{\eta}}^4 + \frac{\beta_2}{2}\big(\eta_x\eta_y^\ast - \eta_y\eta_x^\ast)^2
        + \beta_3\abs{\eta_x\eta_y}^2 \\
        &+\kappa_1\big(\abs{D_x\eta_x}^2 + \abs{D_y\eta_y}^2\big) + \kappa_2\big(\abs{D_y\eta_x}^2 + \abs{D_x\eta_y}^2\big)\\
        &+\kappa_3\big[(D_x\eta_x)^\ast D_y\eta_y + (D_y\eta_y)^\ast D_x\eta_x\big]\\
        &+\kappa_4\big[(D_x\eta_y)^\ast D_y\eta_x + (D_y\eta_x)^\ast D_x\eta_y\big]\\
        &+\kappa_5\big(\abs{D_z\eta_x}^2 + \abs{D_z\eta_y}^2\big) + \frac{\abs{\nabla\times\v{A}}^2}{2\mu_0},
    \end{split}
    \label{eq:GLmodel:fullF_SI_units}
\end{equation}
where $D_a = \nabla_a - i(q/\hbar) A_a$, $q$ is the  charge of the Cooper pair, $\hbar$ is Planck's reduced constant and $\mu_0$ is the vacuum permeability. In this expression, the conventional $xy$-basis is used for the complex fields $\eta_{x}$ and $\eta_y$. 
Rotating this to the chiral basis through the transformation
\begin{equation}
    \eta_\pm = \frac{1}{\sqrt{2}}(\eta_x \pm i \eta_y),
    \label{eq:GLmodel:chiral_transformation}
\end{equation}
yields the energy density
\begin{equation}
    \begin{split}
        \mathcal{F} = &-\alpha\abs{\v{\eta}}^2 + \big(2(\beta_1-\beta_2) + \beta_3\big)\frac{\abs{\eta_+}^4+\abs{\eta_-}^4}{4}\\
        &+ (\beta_1+\beta_2)\abs{\eta_+\eta_-}^2 - \frac{\beta_3}{2}\Re \eta_+^2\eta_-^{\ast 2}\\
        &+\frac{\kappa_1+\kappa_2}{2}\big(\abs{D_x\v{\eta}}^2+\abs{D_y\v{\eta}}^2\big)+ \kappa_5\abs{D_z\v{\eta}}^2 \\
        &+ (\kappa_1-\kappa_2)\Re\big[D_x\eta_+(D_x\eta_-)^\ast - D_y\eta_+(D_y\eta_-)^\ast\big]\\
        &+ (\kappa_4-\kappa_3)\Im\big[D_x\eta_+(D_y\eta_+)^\ast + D_y\eta_-(D_x\eta_-)^\ast\big]\\
        &+ (\kappa_4+\kappa_3)\Im\big[D_x\eta_+(D_y\eta_-)^\ast + D_y\eta_+(D_x\eta_-)^\ast\big]\\
        &+ \frac{\abs{\nabla\times\v{A}}^2}{2\mu_0}.
    \end{split}
    \label{eq:GLmodel:fullF_SI_chiral}
\end{equation}
Taking the mean field limit and looking at the fourth-order terms yields the constraint that for the mean field energy to be bounded from below, then
$\beta_1>0$, $\beta_3>-2\beta_1$ and $\beta_3>2(\beta_2-\beta_1)$.
Minimizing $\mathcal{F}$ w.r.t. $\eta_\pm$ yields the three distinct mean field solutions in Table~\ref{tab:GLmodel:MF_phases}. The regions of the $\beta_3/\beta_1$, $\beta_2/\beta_1$-parameter space for which each of these solutions minimizes $\mathcal{F}$ is shown in Fig.~\ref{fig:GLmodel:MF_phases}. 
One of these solutions, known as the $A$-phase, exhibits spontaneous time-reversal symmetry breaking. This is the phase we are interested in examining.
\begin{table}
    \centering
    \begin{tabular}{cccc}
        Name & $\mathcal{F}$ & $(\eta_+, \eta_-)$ & $u^2$\\\hline\vspace{0.4em}
        A-phase & $-\frac{\alpha^2}{2(\beta_1-\beta_2)+\beta_3}$ & $u(0, 1) \lor u(1,0)$ & $\frac{2\alpha}{2(\beta_1-\beta_2)+\beta_3}$\\
        B-phase & $-\frac{\alpha^2}{2\beta_1+\beta_3}$ & $u(\pm i, 1)$ & $\frac{\alpha}{2\beta_1+\beta_3}$\\
        C-phase & $-\frac{\alpha^2}{2\beta_1}$ & $u(\pm1, 1)$ & $\frac{\alpha}{2\beta_1}$\\\hline
    \end{tabular}
    \caption{Name, mean-field energy density and solution modulo an overall phase, of the mean field minimization of $\mathcal{F}$ in Eq.~\eqref{eq:GLmodel:fullF_SI_chiral}. The $A$-phase is the phase that exhibits spontaneous time-reversal symmetry breaking in zero magnetic field, and is the one we focus on in this paper. The $B$ and $C$-phases are time-reversal symmetric odd-parity superconducting states with line nodes in the gap on the Fermi-surface.}
    \label{tab:GLmodel:MF_phases}
\end{table}

We now focus on the $A$-phase. To write $\mathcal{F}$ on a dimensionless form, we scale the dimensionless  energy density $\tilde{\mathcal{F}}$ and dimensionless fields $\tilde{\eta}_\pm$ by their mean field values in the $A$-phase such that $\mathcal{F} = 2\alpha^2/[2(\beta_1-\beta_2)+\beta_3]\tilde{\mathcal{F}}$ and $\eta_\pm = \sqrt{2\alpha/[2(\beta_1-\beta_2)+\beta_3]}\tilde{\eta}_\pm$. 
We also choose the length scale such that the coefficient in front of the first term in the kinetic part of $\mathcal{F}$ becomes trivial in dimensionless units, i.e. $\nabla_a = \sqrt{2\alpha/(\kappa_1+\kappa_2)}\tilde{\nabla}_a$. Finally we scale the gauge field $A_a = \hbar\sqrt{2\alpha/(\kappa_1+\kappa_2)}/q\tilde{A}_a$ such that $\tilde{D}_a = \tilde{\nabla}_a-i\tilde{A}_a$.
To simplify the notation, we  neglect the tilde in the dimensionless variables in the following. With these choices of units, $\mathcal{F}$ takes the dimensionless form
\begin{equation}
    \begin{split}
        \mathcal{F} = &-\abs{\v{\eta}}^2 + \frac{\abs{\eta_+}^4 + \abs{\eta_-}^4}{2} + \frac{2}{1+\Delta\tilde{\beta}}\abs{\eta_+\eta_-}^2 + \nu\Re\eta_+^2\eta_-^{\ast 2}\\
        &+\abs{D_x\v{\eta}}^2 + \abs{D_y\v{\eta}}^2 + 2\tilde{\kappa}_5\abs{D_z\v{\eta}}^2+ \frac{\abs{\nabla\times\v{A}}^2}{g^2}\\
        &+ (1+\nu_k)\Re\big[D_x\eta_+(D_x\eta_-)^\ast - D_y\eta_+(D_y\eta_-)^\ast\big]\\
        &+ \Delta\tilde{k}\Im\big[D_x\eta_+(D_y\eta_+)^\ast+D_y\eta_-(D_x\eta_-)^\ast\big]\\
        &+ (\nu_k-1-2\Delta)\Im\big[D_x\eta_-(D_y\eta_+)^\ast+D_y\eta_-(D_x\eta_+)^\ast\big],\\
    \end{split}
    \label{eq:GLmodel:full_dim_less_F}
\end{equation}
for dimensionless parameters
\begin{subequations}
\begin{align}
    \Delta\tilde{\beta} &= \frac{\beta_3-3\beta_2+\beta_1}{\beta_1+\beta_2},
    \label{eq:GLmodel:Db_def}\\
    \nu &= \frac{\beta_3}{2(\beta_2-\beta_1)-\beta_3},\label{eq:GLmodel:nu_def}\\
    \nu_k &= \frac{\kappa_1-3\kappa_2}{\kappa_1+\kappa_2},\label{eq:GLmodel:nuk_def}\\
    \Delta\tilde{k} &= 2\frac{\kappa_4-\kappa_3}{\kappa_1+\kappa_2},\label{eq:GLmodel:Dk_def}\\
    \Delta &= \frac{\kappa_3+\kappa_4-2\kappa_2}{\kappa_1+\kappa_2},\label{eq:GLmodel:Delta_def}\\
    \tilde{\kappa}_5 &= \frac{\kappa_5}{\kappa_1+\kappa_2},\label{eq:GLmodel:k5_def}\\
    g &= \frac{q}{\hbar}\sqrt{\mu_0\frac{(\kappa_1+\kappa_2)^2}{2(\beta_1-\beta_2)+\beta_3}}\label{eq:GLmodel:g_def}.
\end{align}
\end{subequations}

So far, no assumptions have been made about the values of the material parameters $\alpha$, $\beta_i$ and $\kappa_i$. 
Based on microscopic derivations of the kinetic constants in the weak coupling and clean limit \cite{Krohg18} we have that $\kappa_2=\kappa_3=\kappa_4\propto\langle v_x^2v_y^2\rangle$ and $\kappa_1\propto\langle v_x^4\rangle$.
For the case of a cylindrical Fermi surface, another microscopic derivation of the Ginzburg-Landau coefficients \cite{Zhu97} shows that in the weak coupling approximation the relations $\beta_2/\beta_1 = \kappa_2/\kappa_1$ and $\beta_3 = 3\beta_2 - \beta_1$ hold.
The validity of these constraints has been extended to non-cylindrical Fermi surfaces in \cite{Agterberg98} and \cite{AgterbergVortex98}.
Using these relationships, we see that $\Delta\tilde{\beta}=\Delta\tilde{k}=\Delta=0$, $g = q/\sqrt{\mu_0\kappa_1^2(1 + \kappa_2/\kappa_1)/\beta_1}/\hbar$ and $\nu=\nu_k$, such that Eq.~\eqref{eq:GLmodel:full_dim_less_F} reduces to Eq.~\eqref{eq:GLmodel:chiral_clean_limit} with the previously mentioned interpretation of $\nu$ as measuring the Fermi surface anisotropy.

The weak coupling relationship $\beta_3 = 3\beta_2-\beta_1$ constrains the system to be along the blue dashed line in parameter space in Fig.~\ref{fig:GLmodel:MF_phases}. This line can in turn be parametrized in terms of $\nu$ such that $\beta_2/\beta_1 = (1-\nu)/(\nu+3)$ and $\beta_3/\beta_1 = -4\nu/(\nu+3)$.
Thus, we see that the Fermi surface is cylindrically symmetric for $\nu=0$, i.e. for $3\beta_2=\beta_1$ where the blue line crosses the $x$-axis.
As $\nu$ approaches $1$, the system approaches the $B$-phase and the Fermi surface becomes completely square.
The Fermi surface also becomes completely square as $\nu$ approaches $-1$, but in this case the system approaches the $C$-phase instead.
\subsection{Lattice Ginzburg Landau model}
\label{sec:GLmodel:Disc}
The  GL energy $E = \int\mathcal{F}\,\mathrm{d}^3r$ in Eq.~\eqref{eq:GLmodel:chiral_clean_limit} is discretized on a $3D$ qubic lattice of points $\v{r}$ containing values for the complex fields $\eta^\pm_\v{r}$ as well as link variables
\begin{equation}
    A_{\v{r},\mu} = \int_{\v{r}}^{\v{r}+l\hat{\mu}}\;A_\mu(\v{r})\;\mathrm{d}r_\mu
    \label{eq:GLmodel:Disc:LinkVarDef}
\end{equation}
between the points at $\v{r}$ and $\v{r} + l\hat{\mu}$, where $l$ is the lattice point separation spacing. On the lattice,  $E = \int\mathcal{F}\,\mathrm{d}^3r$ is written as a lattice sum over the discretized energy density 
\begin{equation}
    E = l^3\sum_{\v{r}} \mathcal{F}^\mathrm{r}, 
\label{eq:GL_Total_Lattice_Energy}
\end{equation}
where $\v{r}$ runs over the sites of the numerical lattice and the lattice energy density  $\mathcal{F}^\mathrm{r} $ is given by
\begin{equation}    
    \mathcal{F}^\mathrm{r} = \mathcal{F}^\mathrm{r}_\text{K} + \mathcal{F}^\mathrm{r}_\text{An} + \mathcal{F}^\mathrm{r}_\text{MG} + \mathcal{F}^\mathrm{r}_\text{V} + \mathcal{F}^\mathrm{r}_\text{A}.
    \label{eq:GLmodel:Disc:F_overview}
\end{equation} 
This defines an effective lattice gauge theory derived from the continuum theory in Eq.~\eqref{eq:GLmodel:chiral_clean_limit}.
In Eq.~\eqref{eq:GLmodel:Disc:F_overview}, $\mathcal{F}^\mathrm{r}$ is split into various gradient terms, a potential-energy term $\mathcal{F}^\mathrm{r}_\text{V}$, and a magnetic field energy density term $\mathcal{F}^\mathrm{r}_\text{A}$, respectively.
The gradient terms have been written as a sum of three different terms to be detailed below, namely a standard isotropic term $\mathcal{F}^\mathrm{r}_\text{K}$, a term contributing to anisotropy in the kinetic energy $\mathcal{F}^\mathrm{r}_\text{An}$, and a mixed gradient term $\mathcal{F}^\mathrm{r}_\text{MG}$.

In the discretized energy density, covariant derivatives are treated by a forward difference scheme
\begin{equation}
    D_\mu\eta^h = (\partial_\mu - iA_\mu)\eta^h \mapsto l^{-1}\big(\eta^h_{\v{r}+l\hat{\mu}}e^{-ilA_{\v{r},\mu}} - \eta^h_\v{r}\big),
    \label{eq:GLmodel:Disc:CovariantDerivative}
\end{equation}
where the field value $\eta^h_{\v{r}+l\hat{\mu}}$ has been parallel transported
back to the point $\v{r}$ via the Abelian $\mathrm{U}(1)$ parallel transporter $U_{\v{r},\mu}=e^{-ilA_{\v{r},\mu}}$ \cite{Munster2000}.
In the following, we set the lattice spacing $l = 1$.

Writing the complex fields $\eta^h_\v{r}$ in terms of their amplitudes $\rho^h_\v{r}$ and phases $\theta^h_\v{r}$, the discretized expression derived from the
kinetic part of $\mathcal{F}$ given by the covariant derivatives in Eq.~\eqref{eq:GLmodel:chiral_clean_limit:kinetic}, is written on the standard cosine form \cite{Galteland15}
\begin{equation}
    \mathcal{F}^\mathrm{r}_\text{K} = 2\sum_{\mu,h}\big[\rho^{h\,2}_\v{r} - \rho^h_\v{r}\rho^h_{\v{r}+\hat{\mu}}\cos\big(\theta^h_{\v{r}+\hat{\mu}} - \theta^h_\v{r} - A_{\v{r},\mu}\big)\big]
    \label{eq:GLmodel:Disc:kinetic}
\end{equation}
Here, $h$ denotes the two chiral components $h\in\{\pm\}$, while $\mu\in\{x,y,z\}$ and we have set the parameter $\tilde{\kappa}_5=1/2$ such as to make the kinetic energy density isotropic. 

Introducing the notation $\bar{h} = -h$, $q\in\{x,y\}$ and the symbol $\zeta_{\alpha \beta} = 1-2\delta_{\alpha \beta}$, the anisotropic part of $\mathcal{F}$ in Eq.~\eqref{eq:GLmodel:chiral_clean_limit:anisotropy} is discretized to
\begin{equation}
    \mathcal{F}^\mathrm{r}_\text{An} = (1+\nu)\sum_{q h}\zeta_{qy}\rho^{\bar{h}}_\v{r}\rho^h_{\v{r}+\hat{q}}\cos\big(\theta^h_{\v{r}+\hat{q}} - \theta^{\bar{h}}_\v{r} - A_{\v{r},q}\big).
    \label{eq:GLmodel:Disc:anisotropy}
\end{equation}
These terms mix the two components, and give different signs of the contributions depending on the direction $\hat{q}$, i.e. anisotropic contributions to the kinetic energy.

The contribution $\mathcal{F}^\mathrm{r}_\text{MG} $ in Eq.~\eqref{eq:GLmodel:Disc:F_overview} is named the mixed gradient terms since these terms mix the gradient directions as well as the different components as seen in Eq.~\eqref{eq:GLmodel:chiral_clean_limit:mgt}. On discretized form, it is given by 
\begin{equation}
    \begin{split}
        \mathcal{F}^\mathrm{r}_\text{MG} = &-(1-\nu)\sum_q\big[\rho^+_\v{r} ~\rho^{-}_\v{r}~\sin(\theta^{-}_\v{r}-\theta^+_\v{r})\\
            &+\sum_h\zeta_{+h} ~ \rho^h_{\v{r}+\hat{q}} ~ \rho^{\bar{h}}_\v{r} ~ \sin\big(\theta^h_{\v{r}+\hat{q}}-\theta^{\bar{h}}_\v{r} - A_{\v{r},q}\big)\\
    &+\rho_{\v{r}+\hat{\bar{q}}}^+ ~ \rho^{-}_{\v{r}+\hat{q}}\sin\big(\theta^{-}_{\v{r}+\hat{q}}-\theta^+_{\v{r}+\hat{\bar{q}}} - (A_{\v{r},q} - A_{\v{r},\bar{q}})\big)\big],
    \end{split}
    \label{eq:GLmodel:Disc:mgt}
\end{equation}
where $\bar{q}\in\{x,y\}\setminus\{q\}$.

The discretized potential part of $\mathcal{F}^\mathrm{r}$ is written as
\begin{equation}
    \begin{split}
         \mathcal{F}^\mathrm{r}_\text{V}  &=  
            (\rho^+_\v{r} \rho^-_\v{r})^2 
            \big( 2 +\nu \cos 2\big(\theta^+_{\v{r}} 
            - \theta^-_{\v{r}} \big) \big)\\
            &+   \sum_h \Big[-( \rho^h_\v{r})^2 + \frac{1}{2}  (\rho^h_\v{r})^4 \Big].
    \end{split}
    \label{eq:GLmodel:Disc:potential}
\end{equation}
The first term in Eq.~\eqref{eq:GLmodel:Disc:potential} originates with 
the term $2|\eta_+\eta_-|^2 + \nu\Re(\eta_+^{\ast 2}\eta_-^2)$ in Eq.~\eqref{eq:GLmodel:chiral_clean_limit:potential}.
Of particular interest in the present context is the factor $\cos 2\big(\theta^+_{\v{r}} 
    - \theta^-_{\v{r}} )$.  This term is minimized for
$2\big( \theta^+_{\v{r}} 
- \theta^-_{\v{r}} \big) = \pi$ for $\nu > 0$, thus potentially locking the phase difference, and breaking the global $U(1)$-invariance of the system associated with the phase-difference $\theta^+_{\v{r}} 
- \theta^-_{\v{r}}$ down to $\mathbb{Z}_2$.
The last line in Eq.~\eqref{eq:GLmodel:Disc:potential} comes from the last term in Eq.~\eqref{eq:GLmodel:chiral_clean_limit:potential}, and represents a soft constraint on the  fluctuations of the amplitude $\rho^h_\v{r}$.

Finally, the gauge field energy is given a non-compact discretization \cite{shimizu12} such that $A_{\v{r},\mu}\in(-\infty,\infty)$ and
\begin{equation}
    \mathcal{F}^\mathrm{r}_\text{A} = g^{-2}(\gv{\Delta}\times\v{A}_\v{r})^2 = g^{-2}\sum_{\mu>\lambda}(\Delta_\mu A_{\v{r},\lambda} - \Delta_\nu A_{\v{r},\lambda})^2,
    \label{eq:GLmodel:Disc:gauge}
\end{equation}
where $\mu,\lambda\in\{x,y,z\}$ and $\Delta_\mu A_{\v{r},\lambda} = A_{\v{r}+\hat{\mu},\lambda} - A_{\v{r},\lambda}$.

The model in Eq.~\eqref{eq:GLmodel:chiral_clean_limit} has thus been formulated on a lattice in terms of two parameters,
namely the coupling constant of the gauge-field to the matter field, $g$, and the parameter $\nu$, describing the anisotropy of the Fermi surface.
We will consider the model in this restricted parameters space to make the problem tractable in Monte-Carlo simulations.
The parameter values $\nu=0.1$ and $g=0.3$ have been used for most of the simulation results presented in this paper.

\subsection{XY basis and pseudo-$\mathbb{CP}^1$-constraint} %
\label{sec:GLmodel:xy}
{ The full Ginzburg-Landau model is still too complex to simulate on a lattice of sufficient size. Therefore, a London approximation is typically used
for this kind of problems (see e.g. \cite{Dasgupta-Halperin,Peskin1978,Kleinert1989gauge,Hove-Mo-Sudbo,Smiseth-Smorgrav-Sudbo,Fossheim-Sudbo-Book}). Taking the London limit
in the chiral $p$-wave case, however, requires special care.
As discussed in detail in \cite{speight2019chiral}, all phase and density degrees of freedom are in general coupled. However, as discussed in the same reference, the mixing between different modes for certain parameters is small, making the
London limit an adequate approximation.
The required conditions that must hold in this study are: (i) the dominant length scale in magnetic field should be much larger than the core size, and (ii) the external field should be sufficiently low so that vortex cores do not overlap. 
Since we are interested primarily in vortex dissociation transition, the 
binding energy comes from mixed gradient terms, which are retained in our approximation. The low temperature configuration we obtain, are consistent with the
solutions found at low temperatures in the full Ginzburg-Landau model \cite{AsleGaraud16}.}

To simplify the model, we introduce a pseudo-$\mathbb{CP}^1$-constraint on the complex fields $\eta^\pm_\v{r}$. Since these fields are related to corresponding $xy$-basis
fields $\eta^a_\v{r}$ for $a\in\{x,y\}$ through the orthonormal transformation in Eq.~\eqref{eq:GLmodel:chiral_transformation} we may rotate the expressions
for the discretized free energy densities in Eq.~\eqref{eq:GLmodel:Disc:potential}, \eqref{eq:GLmodel:Disc:kinetic}, \eqref{eq:GLmodel:Disc:anisotropy} and \eqref{eq:GLmodel:Disc:mgt} back to this basis.
It is this $xy$-basis that is used in all simulations when evaluating the free energy for accepting new states through the Metropolis-Hastings algorithm,
since, as we shall see, this ensures that mixed component terms are retained in the London-limit.

The conventional kinetic energy contribution in Eq.~\eqref{eq:GLmodel:Disc:kinetic} is invariant under the change of basis, such that
\begin{equation}
    \mathcal{F}^\mathrm{r}_\text{K} = 2\sum_{a\mu}\big[\rho^{a\,2}_\v{r} - \rho^a_{\v{r}+\hat{\mu}}\rho^a_\v{r}\cos\big(\theta^a_{\v{r}+\hat{\mu}} - \theta^a_\v{r} - A_{\v{r},\mu}\big)\big].
    \label{eq:GLmodel:xy:kinetic}
\end{equation}
The expression for the on-site potential terms however, becomes slightly more involved, perhaps most succinctly expressed as
\begin{equation}
    \begin{split}
        \mathcal{F}^\mathrm{r}_\text{V} = &(1+\nu)\frac{\rho^{x\,4}_\v{r}+\rho^{y\,4}_\v{r}}{4} + \sum_a\big[-\rho^{a\,2}_\v{r} + \frac{1}{2}\rho^{a\,4}_\v{r}\big]\\
        + &(1-\nu)(\rho^x_\v{r}\rho^y_\v{r})^2\big[1 + \frac{1}{2}\cos 2(\theta^x_\v{r}-\theta^y_\v{r})\big].
    \end{split}
    \label{eq:GLmodel:xy:potential}
\end{equation}
The anisotropy-term remains similar in both basis, with the $xy$-basis version having the form
\begin{equation}
    \mathcal{F}^\mathrm{r}_\text{An} = (1+\nu)\sum_{aq}\zeta_{aq}\rho_{\v{r}+\hat{q}}^a\rho^a_\v{r}\cos\big(\theta^a_{\v{r}+\hat{q}} - \theta^a_\v{r} - A_{\v{r},q}\big).
    \label{eq:GLmodel:xy:anisotropy}
\end{equation}
The minor difference being that $\zeta_{aq}$ now depends on both summation indices. Finally, the mixed-gradient terms take the form
\begin{equation}
    \begin{split}
        \mathcal{F}^\mathrm{r}_\text{MG} = &(1-\nu)\sum_a\Big[\rho^a_\v{r}\rho^{\bar{a}}_\v{r}\cos\big(\theta^a_\v{r}-\theta^{\bar{a}}_\v{r}\big)\\
            &-\sum_q\rho^a_{\v{r}+\hat{q}}\rho^{\bar{a}}_\v{r}\cos\big(\theta^a_{\v{r}+\hat{q}} - \theta^{\bar{a}}_\v{r} - A_{\v{r},q}\big)\\
        &+\rho^a_{\v{r}+\hat{x}}\rho^{\bar{a}}_{\v{r}+\hat{y}}\cos\big(\theta^a_{\v{r}+\hat{x}} - \theta^{\bar{a}}_{\v{r}+\hat{y}} - (A_{\v{r},x}-A_{\v{r},y})\big)\Big].
    \end{split}
    \label{eq:GLmodel:xy:mgt}
\end{equation}
The process of discretization commutes with the basis rotation, i.e. first rotating the basis in Eq.~\eqref{eq:GLmodel:chiral_clean_limit} and then discretizing the result yields the same expressions for $\mathcal{F}^\mathrm{r}$.

The model is now simplified by taking the London-limit in the $xy$-basis, i.e. neglecting $xy$-basis amplitude fluctuations such that $\rho^a_\v{r} = \rho^a \forall \v{r}$.
The mean field $A$-phase solution of Eq.~\eqref{eq:GLmodel:chiral_clean_limit} in the $xy$ basis gives amplitudes $\rho^x = \rho^y = 1/\sqrt{2}$ which will be used in the following. Using the $xy$-basis has the comparative advantage over the chiral-basis in that setting the London-limit amplitudes equal to the mean field solution amplitude values does not eliminate the mixed component terms. 
Taking the limit in the $xy$ basis allows the chiral basis amplitudes to fluctuate since from Eq.~\eqref{eq:GLmodel:chiral_transformation} they
are related to their $xy$ counterparts through
\begin{equation}
    \rho^{\pm\,2}_\v{r} = \frac{\rho^{x\,2} + \rho^{y\,2}}{2} \pm \rho^x\rho^y\sin\big(\theta^x_\v{r}-\theta^y_\v{r}).
    \label{eq:GLmodel:xy:chiralAmplitudes}
\end{equation}
From this equation, we see that the $xy$ basis London limit implies the restriction
\begin{equation}
    \rho_\v{r}^{+\,2} + \rho_\v{r}^{-\,2} = \rho^{x\,2} + \rho^{y\,2} = 1,
    \label{eq:GLmodel:xy:CP1}
\end{equation}
and in this sense the London-limit in the $xy$-basis may equivalently be viewed as a $\mathbb{CP}^1$ constraint on the chiral amplitudes $\rho_\v{r}^h$.
Note that a phase-locking of $\theta^x_\v{r}-\theta^y_\v{r} = \pm \pi/2$ corresponds to spontaneous time-reversal symmetry-breaking in zero magnetic field, i.e. $|\eta^+_\v{r}|^2 \neq |\eta^-_\v{r}|^2$.  

Since the $xy$-basis London limit removes two real degrees of freedom from the problem, we expect two constraints in the chiral basis as well.
The second constraint takes the form of the relationship
\begin{equation}
    \tan\theta_\v{r}^+ = \tan\Big(\theta^-_\v{r}+\frac{\pi}{2}\Big)
    \label{eq:GLmodel:xy:chiralPhasesConstraint}
\end{equation}
between the chiral phases. A derivation of this relationship can be found in Appendix~\ref{app:chiralPhases}. This implies that $\theta^+ = \theta^- + \pi/2 + \pi n$ for $n\in\{-2, -1, 0, 1\}$ since phases are defined compactly by $\theta\in[-\pi, \pi)$.
That the phases are not completely locked to each other allows there to be a vortex singularity in one component independent of the other.

\subsection{Symmetrization and lattice potential}

The discretization procedure in Eq.~\eqref{eq:GLmodel:Disc:CovariantDerivative} does not in general guarantee that the resulting discrete lattice free energy is symmetric
under the same transformations as the original continuum theory. It only guarantees that the continuum limit of the discrete theory satisfies these symmetries. To ensure that the lattice energy density is invariant under a four-fold rotation of the numerical lattice, we introduce a  
symmetrization of the discretized $xy$-basis free energy density as follows
\begin{equation}
    \mathcal{F}^\mathrm{s} = \frac{1}{4}\Big[\mathcal{F}^\mathrm{r} + C_4\mathcal{F}^\mathrm{r} + C_4^2\mathcal{F}^\mathrm{r} + C_4^3\mathcal{F}^\mathrm{r}\Big],
    \label{eq:GLmodel:symm:freeEn}
\end{equation}
where $C_4$ is a counter-clockwise rotation by $\pi/2$ radians about the $\hat{z}$ axis,
and we allow lattice translations because of periodic boundary conditions (see next Section).

Under this rotation, we let the gauge-field link-variables $A_{\v{r},\mu}$ transform as the components of a vector field such that
\begin{equation}
    C_4 : A_{\v{r},\mu} = A_{C_4\v{r},C_4\mu}.
    \label{eq:GLmodel:symm:Atransform}
\end{equation}
Since link-variables are only defined for positive directions from any numerical lattice point $\v{r}$, we use the relationship
$A_{\v{r},-\mu} = -A_{\v{r}-\hat{\mu},\mu}$
whenever the transformation in Eq.~\eqref{eq:GLmodel:symm:Atransform} results in a negative link direction.
As a non-trivial example $C_4 : A_{\v{r}+\hat{x},y} = -A_{C_4\v{r}+\hat{y}-\hat{x},x}$.

To figure out how the complex fields $\eta_a$ transform, we remember that they are the coefficients of the vector $\v{d} = \sum_a\eta_a\v{b}_a$ whose
basis vectors $\{\v{b}_a\}$ transform according to the irreducible representation $\Gamma_{5u}$ \cite{Krohg18}.
Inserting the $C_4$ representation matrix then yields the transformation
\begin{equation}
    C_4 :
    \begin{pmatrix}
        \eta^x_\v{r}\\
        \eta^y_\v{r}
    \end{pmatrix}
    =
    \begin{pmatrix}
        0 & -1\\
        1 & 0
    \end{pmatrix}
    \begin{pmatrix}
        \eta^x_{C_4\v{r}}\\
        \eta^y_{C_4\v{r}}
    \end{pmatrix}
    =
    \begin{pmatrix}
        -\eta^y_{C_4\v{r}}\\
        \eta^x_{C_4\v{r}}
    \end{pmatrix}.
    \label{eq:GLmodel:symm:etaTrans}
\end{equation}

Inserting this transformation into the continuous free energy density $\mathcal{F}$ in Eq.~\eqref{eq:GLmodel:fullF_SI_units}, and remembering to also transform the covariant gradients,
it is readily verified that all terms are invariant under $C_4$ as indeed they need to be since $C_4\in D_{4h}$.
For the discretized $xy$-basis version of the same free energy in Eqs.~\eqref{eq:GLmodel:Disc:kinetic}-\eqref{eq:GLmodel:Disc:gauge}, it is then
similarly possible to check that all terms are invariant under $C_4$ except for the mixed gradient terms in Eq.~\eqref{eq:GLmodel:xy:mgt}.
The reason why this term is not symmetric, but the continuum version is, is again that the forward difference discretization procedure
in Eq.~\eqref{eq:GLmodel:Disc:CovariantDerivative} introduces artificial
anisotopies in the system; usually referred to as lattice potentials and does not in general guarantee that the discretized version satisfies all
continuum symmetries. In this particular case, it manifests as an explicit asymmetry because the gradients are in different directions in the same term.

Since all other terms than the mixed gradient terms are already symmetric w.r.t. $C_4$, it suffices to only present the rotated version of this particular term when calculating the symmetrized lattice free energy density $\mathcal{F}^s$.
The details of this expression can be found in Appendix~\ref{app:symm_MG}.

\subsection{Boundary conditions and Landau gauge}

The gauge field link variables are split into a fluctuating- and a constant part such that $A_{\v{r},\mu} = A^f_{\v{r},\mu} + A^c_{\v{r},\mu}$.
Periodic boundary conditions are used in the fluctuating part $A^f_{\v{r},\mu}$, as well as in the discretized field components such that $\eta^a_\v{r} = \eta^a_{\v{r}+L_\mu}$.
For the constant part $A^c_{\v{r},\mu}$, twisted boundary conditions are used  by employing the extended Landau gauge forcing a fixed magnetic flux through the system. 
The extended Landau gauge is given by 
\begin{equation}
    A_{\v{r},q}^c = \frac{2\pi m_q}{L_{\bar{q}}}r_{\bar{q}},\quad A_{\v{r},z}^c = 0,
    \label{eq:GLmodel:gauge:linkVariables}
\end{equation}
where $m_q\in\mathbb{Z}$ and the conditions in \cite{Nguyen99PRB, Nguyen99EPL} have already been incorporated.
This definition makes the full link variable boundary conditions periodic modulo $2\pi$, which prevents geometric frustration.
Eq.~\eqref{eq:GLmodel:gauge:linkVariables} together with periodic boundary conditions for $A^f_{\v{r},\mu}$ forces the system to satisfy the property
$\oint \v{A}\cdot\mathrm{d}\v{r}_\perp = 2\pi fL_xL_y$. This gives a magnetic flux $\v{B} = 2\pi f\hat{z}$ through the system for filling fraction
\begin{equation}
    f = \frac{m_y}{L_x} - \frac{m_x}{L_y}.
    \label{eq:GLmodel:gauge:f}
\end{equation}
The filling fraction then gives the number of magnetic field vortex quanta pr. plaquette in the $xy$-plane.
In the results presented in this paper the choice $m_y = 1, m_x = 0$, 
which reduces the gauge to the normal Landau gauge,
has been used for a system with $L_x=L_y=64$ which yields $f = 1/64$. The qualitative conclusions have however been tested for the symmetric choice
$m_y = 1, m_x = -1$. This choice is symmetric in the sense that in this case we may write $\v{A}^c = -\v{r}\times\v{B}/2$ for $\v{B} = 4\pi/L\hat{z}$.

\section{Details of the numerical calculations} \label{sec:mc}

\subsection{Monte-Carlo update method}

For the Monte-Carlo simulations, the Metropolis-Hastings method \cite{Hastings70} was used to sample states with a probability distribution given by the free energy in Eq.~\eqref{eq:GLmodel:Disc:F_overview}.
This method fulfills the detailed-balance criteria such that importance sampling gives thermodynamic averages as simple
arithmetic averages over the sampled states \cite{Katzgraber09, Press07, Newman99}. 
This method, as well as all other numerics, was implemented in the Julia programming language \cite{Julia17} version $1.0.3$.

As described in Section~\ref{sec:GLmodel:Disc}, the free energy was discretized on a cubic lattice of size $L_x\times L_y\times L_z$. Each lattice point contains one fluctuating variable for each of the $xy$-basis phases: $\theta^x_\v{r}$ and $\theta^y_\v{r}$, and three fluctuating link variables for the gauge field, one for each direction of space: $A_{\v{r},x}$, $A_{\v{r},y}$ and $A_{\v{r},z}$. 
A Monte-Carlo update consists in this case of proposing new values of all these variables, which proposes a new state of a single lattice point and then rejecting or accepting this state according to the Metropolis-Hastings method. A Monte-Carlo sweep then consists of doing this for each individual lattice point.
New values of the phases were proposed uniformly on an open interval $\theta^x_\v{r},\theta^y_\v{r}\in[0, 2\pi)$ using the Julia \lstinline{rand()} function which uses the Mersenne-Twister algorithm \cite{Matsumoto98}. 
The gauge-field link-variables were updated by a uniformly distributed random value $A_{\v{r},q}'$ in a symmetric region centred on the previous value $A_{\v{r},q}$, such that $A_{\v{r},q}'-A_{\v{r},q}\in[-A,A]$.
The parameter $A$ which sets the size of the region, was set to $A=0.1$ based on the fact that at this value at high temperature,
the percentage of proposed states that were accepted was $\sim30\%$.

In order to facilitate efficient computation on highly parallelized computer systems, the numerical lattice was divided into sub-lattices that communicated with each-other as their lattice points were updated. 
The number of sub-lattices was chosen according to what gave the fastest average performance of Monte-Carlo sweeps, which for cubic systems of size $L=64$ turned out to be $16$ sub-lattices. A single MC-sweep was then performed in, on average, $0.11\pm0.01$s.

\subsection{Observables}
To study the model in Eq.~\eqref{eq:GLmodel:chiral_clean_limit} in the chiral basis, the $xy$-basis variables were converted into
their chiral counterparts through Eq.~\eqref{eq:GLmodel:chiral_transformation}. Since the trigonometric formulas for obtaining
the chiral phases $\theta^h_\v{r}$ diverges when $|\eta_\pm|\to0$,
these were expanded to $4$th order to handle this case. The technical details
of this can be found in Appendix~\ref{app:phaseConversion}

{ To characterize the  vortices,
we calculate the curl of the gauge-invariant phase-difference of each chiral component, namely 
$(\v{\nabla}\times(\v{\nabla}\theta^h-\v{A}))/2\pi$. 
This amounts to calculating the 
 lattice curl of the gauge-invariant phase-difference $\Delta_q\theta^h_\v{r}-A_{\v{r},q}$ around a fundamental
plaquette of the numerical lattice.  By adding the constant magnetic flux density $f$, we obtain a quantity which we will call the local vorticity of each component \cite{shimizu12}
\begin{equation}
    n^h_{\v{r},z} = \frac{1}{2\pi}\epsilon_{zij}\Delta_i\big(\Delta_j\theta^h_{\v{r}} - A_{\v{r},j}\big)_\pi + f,
    \label{eq:GLmodel:Obs:nz}
\end{equation}
where implicit summation over indices is understood and $\epsilon_{zij}$ is the Levi-Civita symbol.
$\big(\phi\big)_\pi$ is shorthand notation for $\mod(\phi+\pi, 2\pi)-\pi$, 
which draws the argument back into the primary interval $[-\pi,\pi)$. 
The filling fraction $f$ is defined in Eq.~\eqref{eq:GLmodel:gauge:f} and gives the number of fundamental vortex quanta pr. planar plaquette
as determined by the extended Landau gauge \cite{Li93,shimizu12,Kragset08}.
Note that 
$\Delta_q\theta^h_\v{r}-A_{\v{r},q}$
in general does not give the current of each component in the $p$-wave case, but is sufficient to distinguish the structure of vortices and to compare with the results in \cite{AsleGaraud16}.}

The $z$-averaged vorticity is then naturally defined as
\begin{equation}
    n^h_{\v{r}_\perp,z} = \frac{1}{L_z}\sum_{r_z=0}^{L_z-1}n^h_{\v{r},z},
    \label{eq:GLmodel:Obs:zAvg}
\end{equation}
which is used through its thermal average $\langle n^h_{\v{r}_\perp,z}\rangle$ in order to obtain detailed information about the real space structure
of the vortex lattices as well as of the vortex cores in the present model.

A related observable is the $z$-averaged gauge invariant chiral phase difference
\begin{equation}
    \delta\theta_{\v{r}_\perp} = \Big\langle\frac{1}{L_z}\sum_{r_z=0}^{L_z-1}\big(\theta^+_{\v{r}} - \theta^-_{\v{r}}\big)_\pi\Big\rangle,
    \label{eq:GLmodel:Obs:phaseDiffAvg}
\end{equation}
where $\langle\cdot\rangle$ denotes thermal averaging. This observable is also useful in studying the nature of the vortices.

To extract a clearer picture of the overall spatial correlations of the vortex lattice we define the structure function
\begin{equation}
    S^h(\v{k}_\perp) = \frac{1}{(fL_xL_y)^2}\Big\langle \Big| \sum_{\v{r}_\perp}n^h_{\v{r}_\perp,z}e^{i\v{k}_\perp\cdot\v{r}_\perp} \Big|^2 \Big\rangle,
    \label{eq:GLmodel:Obs:structureFunction}
\end{equation}
which essentially amounts to taking the planar Fourier transform of the $z$-averaged vorticity.
The fast-Fourier algorithm was used to efficiently compute the structure function for all Bragg-points $\v{k}_\perp$.
The structure function is normalized such that $S^h(\v{0}) = 1$.

For any vortex lattice signature, the structure function is expected to exhibit peaks at characteristic Bragg points situated equidistantly from the origin.
For a hexagonal lattice we expect $6$ peaks with $\pi/3$ mutual angular distance,
while for a square lattice we expect $4$ peaks with $\pi/2$ mutual angular distance.
To distinguish these two signals clearly, the histogram
\begin{equation}
    h(\delta\Delta\phi^h) = \frac{1}{\delta\Delta\phi^h |\{\Delta\phi_i^h\}|}\sum_{\{\Delta\phi^h_i\}}\delta_{\Delta\phi^h_i\in\delta\Delta\phi^h},
    \label{eq:GLmodel:Obs:hist}
\end{equation}
is constructed, where $\delta\Delta\phi^h$ is some angular interval bin. 
The angular distances $\Delta\phi^h_i$ are obtained by calculating $S^h(\v{k}_\perp)$ using a certain number of Monte-Carlo measurements, then
finding the radius $\abs{\v{k}_\perp}_m$
which yields the largest value of $\int_0^{2\pi}S^h(\abs{\v{k}_\perp},\phi)\mathrm{d}\phi$.
A ribbon is then constructed around this radius from which the largest value of $S^h(\v{k}_\perp)$ is picked for each angle such that
\begin{equation}
    S^h(\phi) = \max\big\{S^h(\abs{\v{k}_\perp},\phi)\;:\;\abs{\abs{\v{k}_\perp} - \abs{\v{k}_\perp}_m} < k_r\big\}.
    \label{eq:GLmodel:Obs:polarStruct}
\end{equation}
The angular positions $\{\phi_i\}$ of the $6$ highest peaks in $S^h(\phi)$ are then found.
Finally all mutual distances between these positions are found $\{\Delta\phi_i\} = \{\abs{\phi_k - \phi_j}\;:\;\phi_k<\phi_j\}$ which is used to calculate the histogram $h(\delta\Delta\phi^h)$.
The process is repeated for independent Monte-Carlo measurements of $S^h(\v{k}_\perp)$ until there are sufficient $\Delta\phi_i$
to construct the histogram.

The above quantities, taken together, provide considerable information on not only the symmetry of the vortex lattices at various temperatures,
but also on the structure of the vortex cores corresponding to each lattice symmetry.

The critical temperature at the position of the upper critical field crossover-line $H_{c2}(T)$ was found by examining the specific heat
\begin{equation}
C_v = \beta^2 \big( \langle E^2 \rangle -\langle E \rangle^2 \big),
\end{equation}
and a chiral order parameter that we will now describe.
The Higgs field components $\eta^+$ and $\eta^-$ are related through the time-reversal transformation, hence a difference in their density
signify a spontaneous breaking of $\mathbb{Z}_2$ time-reversal symmetry.
Since this density can be measured by the component amplitudes, a
useful chiral order parameter is given by
\begin{equation}
    \begin{split}
        \delta u^2 = &\Big|\Big\langle\frac{1}{L_xL_yL_z}\sum_\v{r}\big(\rho^{+\,2}_\v{r} - \rho^{-\,2}_\v{r}\big)\Big\rangle\Big|\\
        = &\frac{2\rho^x\rho^y}{L_xL_yL_y}\Big|\Big\langle\sum_\v{r}\sin(\theta^x_\v{r}-\theta^y_\v{r})\Big\rangle\Big|.
    \end{split}
    \label{eq:GLmodel:Obs:du2}
\end{equation}
From the last line it is clear that it is the locking of the $xy$ phase-difference that is responsible for the breaking of time-reversal symmetry.

\subsection{Thermalization and measurement steps}
Before measurements of observables were performed, the lattice was initialized with random values for all fluctuating variables at each
lattice point,  resulting in high energy states. Then, a two-step thermalization procedure was done which consisted of a stepwise decrease in
temperature to decrease the chance of a meta-stable state, followed by a number of basic Monte-Carlo sweeps (thermalization). The steps during the cooldown procedure were distributed as
a geometric series between a high- and low-temperature, so that more MC-sweeps would be concentrated at lower temperatures.
During cooldown, $\approx1.3 \cdot 10^5$ MC-sweeps were distributed equally on $1024$ temperature steps. This was then followed by $\approx1.3 \cdot 10^5$ additional
MC-sweeps that were discarded before measurements began. 
To confirm that this yielded a properly thermalized state, we checked that the internal  energy of the system as a function of MC-time had converged and remained stable during measurements.

$256$ intermediate MC-sweeps were performed between each measurement to diminish auto-correlation effects. 
The number of measurements of observables varied between simulations, from $1024$ for sampling at high temperatures, to $4096$ when estimating $C_v$ close to the phase transition.

Measurements were performed sequentially by lowering the  temperature, such that the last state of the lattice in the measurement series at one temperature, was used as the initial state when thermalizing the simulation for the next lower temperature. To prevent the simulation from getting stuck in a meta-stable state, several series of simulations were performed using independent initial states to verify the validity of the results.

\subsection{Post-processing}
Multi-histogram Ferrenberg-Swendsen  reweighting \cite{FS_1988,FS_1989} was used to calculate the specific heat $C_v$ accurately at temperatures close the peak in $C_v$. The non-linear Ferrenberg-Swendsen equations for free energy were solved self-consistently and iteratively using the Julia NLsolve library in which automatic forward differentiation was used to find the Jacobian and a trust-region method was used as the iterative algorithm \cite{Nocedal06}.

The jackknife method 
\cite{Efron_1_1979} and Ferrenberg-Swendsen reweighting \cite{FS_1988,FS_1989}
were used to compute averages and uncertainties of observables. 
The number of blocks dividing  the  measurement series in the jackknife method  was  set  to  $128$.  This  gave a  block  length  where the estimate of variance had leveled off, indicating that autocorrelations had  effectively been reduced. 

\section{Lattice Structures}    \label{sec:latticeStructures}
\begin{figure}[h]
    \subfloat[\label{fig:schematic:1q:v}]{%
        \includegraphics[width=.23\textwidth]{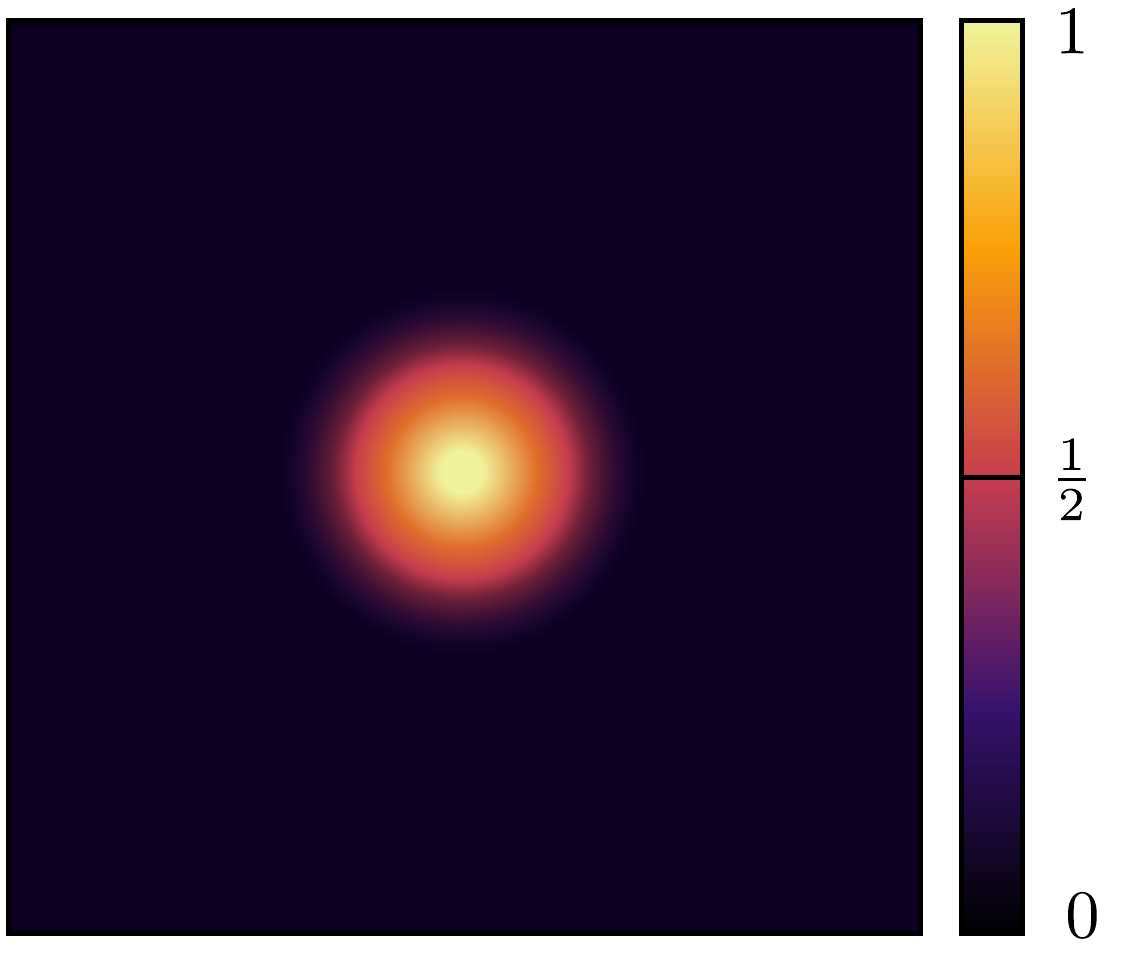}
    }\hfill
    \subfloat[\label{fig:schematic:2q:v}]{%
        \includegraphics[width=.23\textwidth]{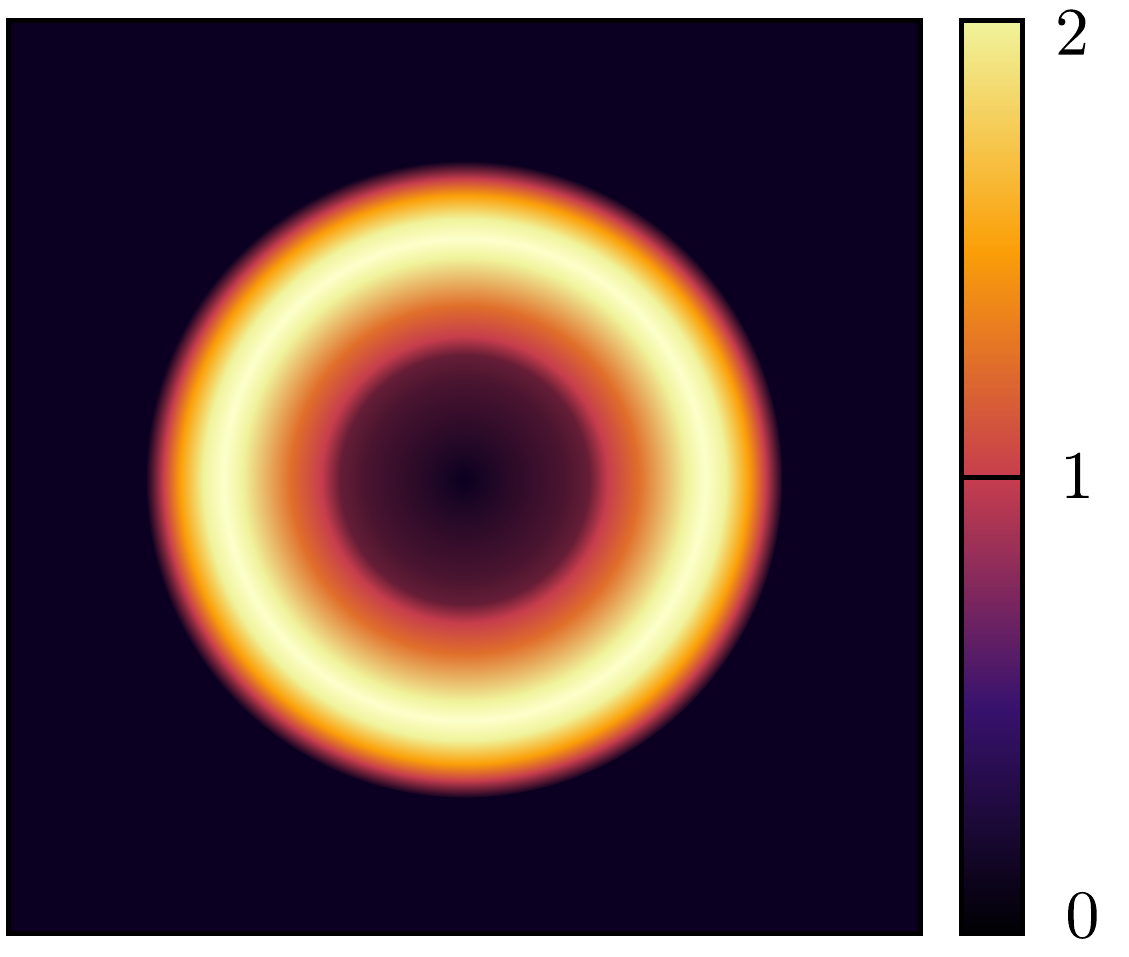}
    }

    \subfloat[\label{fig:schematic:1q:pd}]{%
        \includegraphics[width=.23\textwidth]{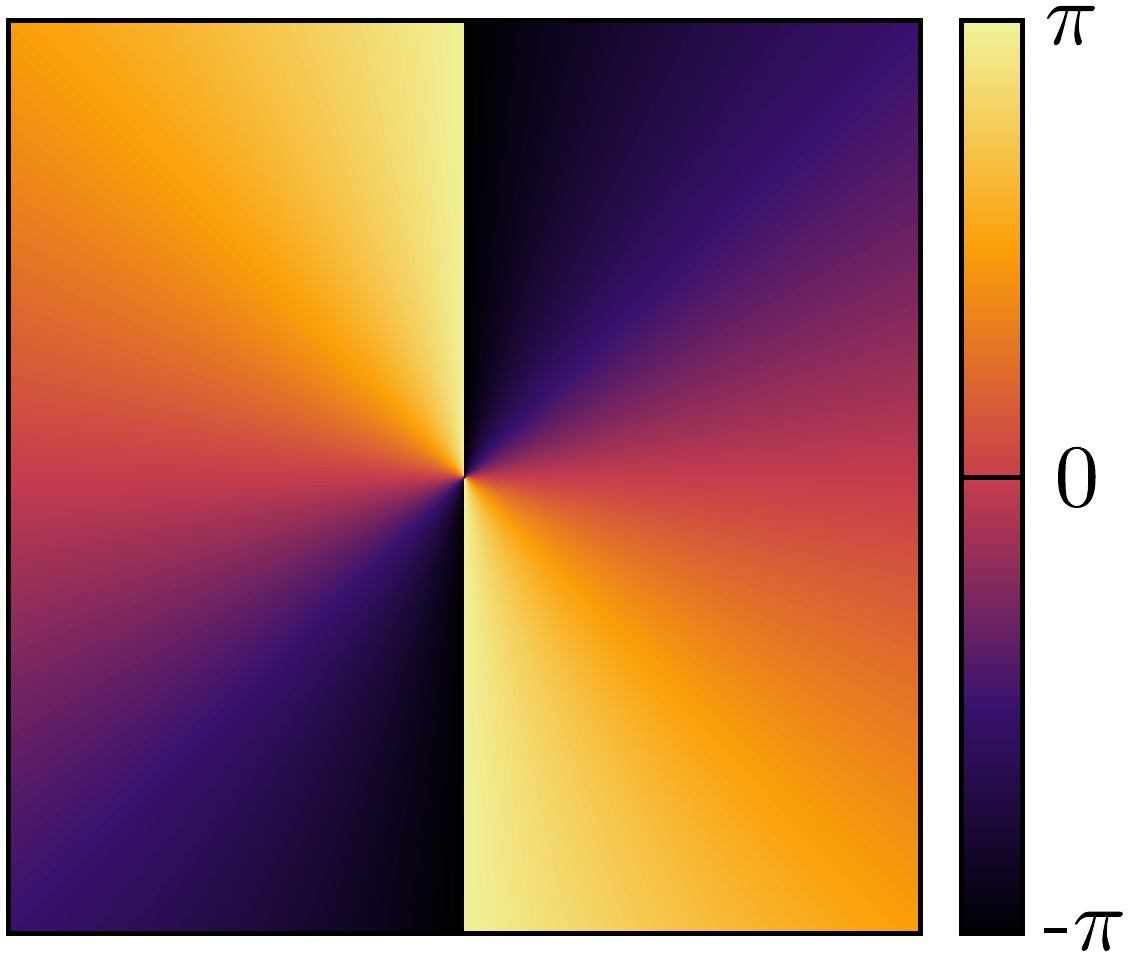}
    }\hfill
    \subfloat[\label{fig:schematic:2q:pd}]{%
        \includegraphics[width=.23\textwidth]{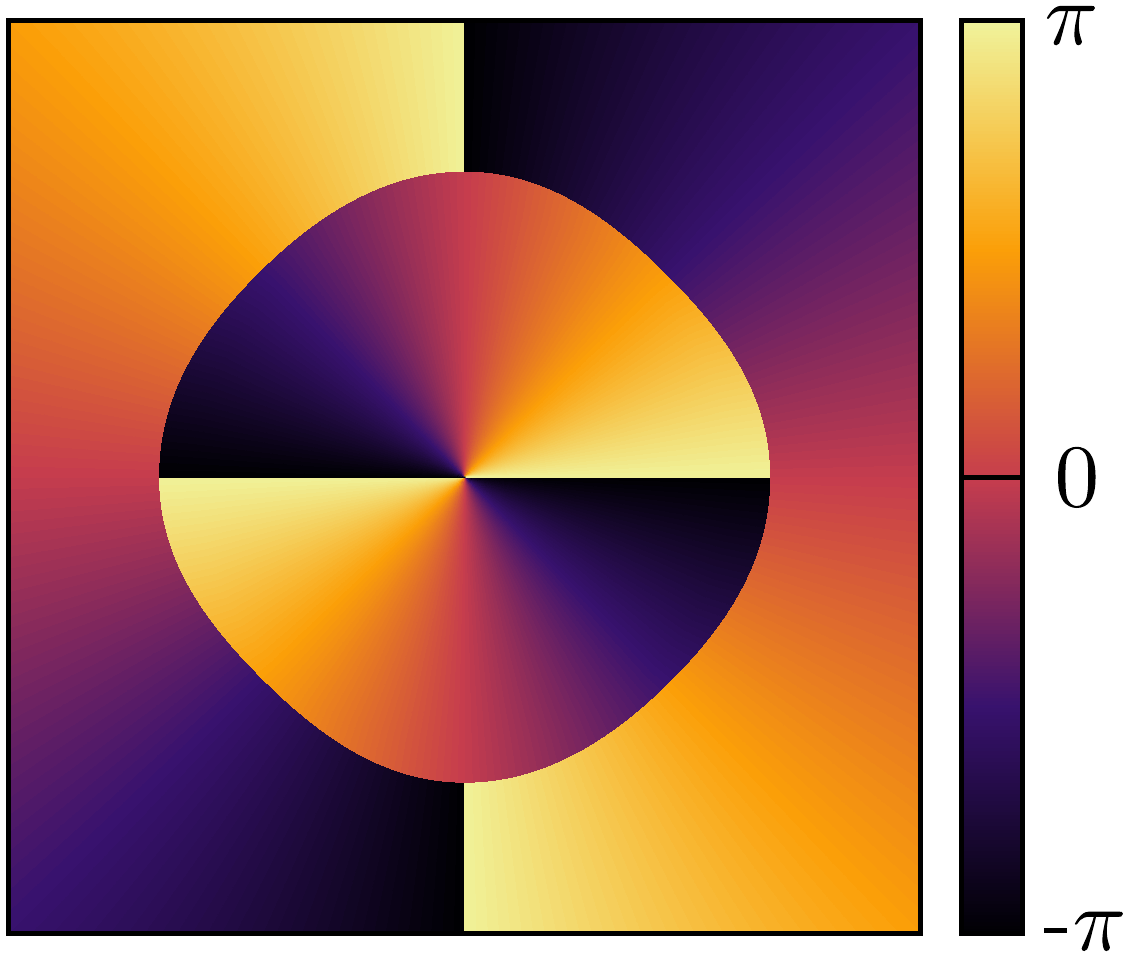}
    }
    \caption{Schematic of vorticities and corresponding  phase difference signature $\theta^+-\theta^-$ of vortices for $H>0$. 
     \protect\subref{fig:schematic:1q:v} and \protect\subref{fig:schematic:1q:pd} shows vorticity and phase-difference respectively for a 
     singly-quantized vortex with winding number $n_+=1$ and $n_-=-1$.
     \protect\subref{fig:schematic:2q:v} and \protect\subref{fig:schematic:2q:pd} shows vorticity and phase difference respectively for a doubly-quantized vortex with winding number $n_+=2$ and $n_-=0$. The figures are directly based on the ones presented in Ref.~\cite{AsleGaraud16}.}
    \label{fig:schematic}
\end{figure}

\captionsetup[subfloat]{position=top}
Before we present our numerical results based on our large-scale Monte-Carlo simulations, we provide a schematic introduction to the results, to assist the reader. In Fig.~\ref{fig:schematic}, we show schematically vorticities and phase-windings that we expect to find for two different types of vortices. In the following, the notation will be as follows.
A phase winding in chiral component $\eta_+$ of $2 \pi n_+$ and in chiral component $\eta_-$ of $2 \pi n_-$ will be denoted $(n_+,n_-)$. 
A vortex with $(n_+,n_-)= (1,-1)$ will be denoted as singly-quantized. 
A vortex with $(n_+,n_-)= (2,0)$ will be denoted as doubly-quantized. 

For a singly-quantized vortex, the vorticity is expected to have a magnetic field-profile centered at the origin, with a maximum magnetic field at the origin, see Fig.~\ref{fig:schematic:1q:v}. The corresponding phase-winding is shown in Fig.~\ref{fig:schematic:1q:pd}. Note the four-fold symmetry in the color-pattern, the radial monotonicity in the phase-value away from the origin, and the $2\pi$-discontinuity along the horizontal axis. For a doubly-quantized vortex, the vorticity is expected to have a magnetic field-profile centered at the origin, with a minimum magnetic field at the origin and a ring of maxima away from the origin, see Fig.~\ref{fig:schematic:2q:v}. The corresponding phase-winding is shown in Fig.~\ref{fig:schematic:2q:pd}. The main difference from the phase winding in Fig.~\ref{fig:schematic:1q:pd}, is the inner circle close to the origin, where phase-windings are rotated by $\pi/2$ compared to the phase-windings in Fig.~\ref{fig:schematic:1q:pd}. For a detailed discussion of this point, see also Section III of \cite{AsleGaraud16}. 

\newlength{\tableauWidth}
\setlength{\tableauWidth}{0.45\textwidth}
\newlength{\markedPltWidth}
\setlength{\markedPltWidth}{0.36\textwidth}

This will be our main diagnostic tool for identifying whether vortices are singly or doubly quantized. As a check on this, we will count the total vorticity in each component and check that this corresponds to the total vorticity of the system, given by the external magnetic field. 

In the following, we focus on results obtained for the parameter set $\nu=0.1,g=0.3,f=1/64$. 
The parameter $\nu=0.1$ corresponds to a moderately four-fold anisotropic Fermi-surface. 
To set the temperature-scale of our finite-field simulations, we have found it useful to first perform Monte-Carlo simulations in zero field, to locate the maximum of the specific heat $C_v$. 
This maximum occurs at $T \approx 2.016\pm0.002$ for $f=0$, which we denote as the critical temperature $T_c$ of the superconductor. 
A rounded and suppressed peak in the specific heat persists at $f>0$. For $f=1/64$, this rounded peak (no longer a phase transition) occurs at $T = 1.86\pm0.04$. 
$T=1.86$ is therefore a natural temperature-scale for the vortex system at $f=1/64$. For this filling fraction, we only expect to see vortex lattice structures for $T < 1.86$.

We will mainly present results starting with high temperatures and then proceeding to lower temperatures. At high temperatures, we will find a plasma phase totally dominated by thermally induced vortex-loops. Proceeding to lower temperatures where a vortex-lattice forms, we find a singly-quantized square vortex lattice. Lowering the temperatures further, we eventually find a doubly-quantized hexagonal lattice. At the end, we briefly discuss a "mixed" phase of singly-quantized and doubly quantized vortices, located at intermediate temperature between the doubly-quantized and singly-quantized vortex lattice phases.

\subsection{Specific heat and chiral order parameter}\label{subsec:f=0}
\begin{figure}[h]
    \centering
    \includegraphics[width=0.5\textwidth]{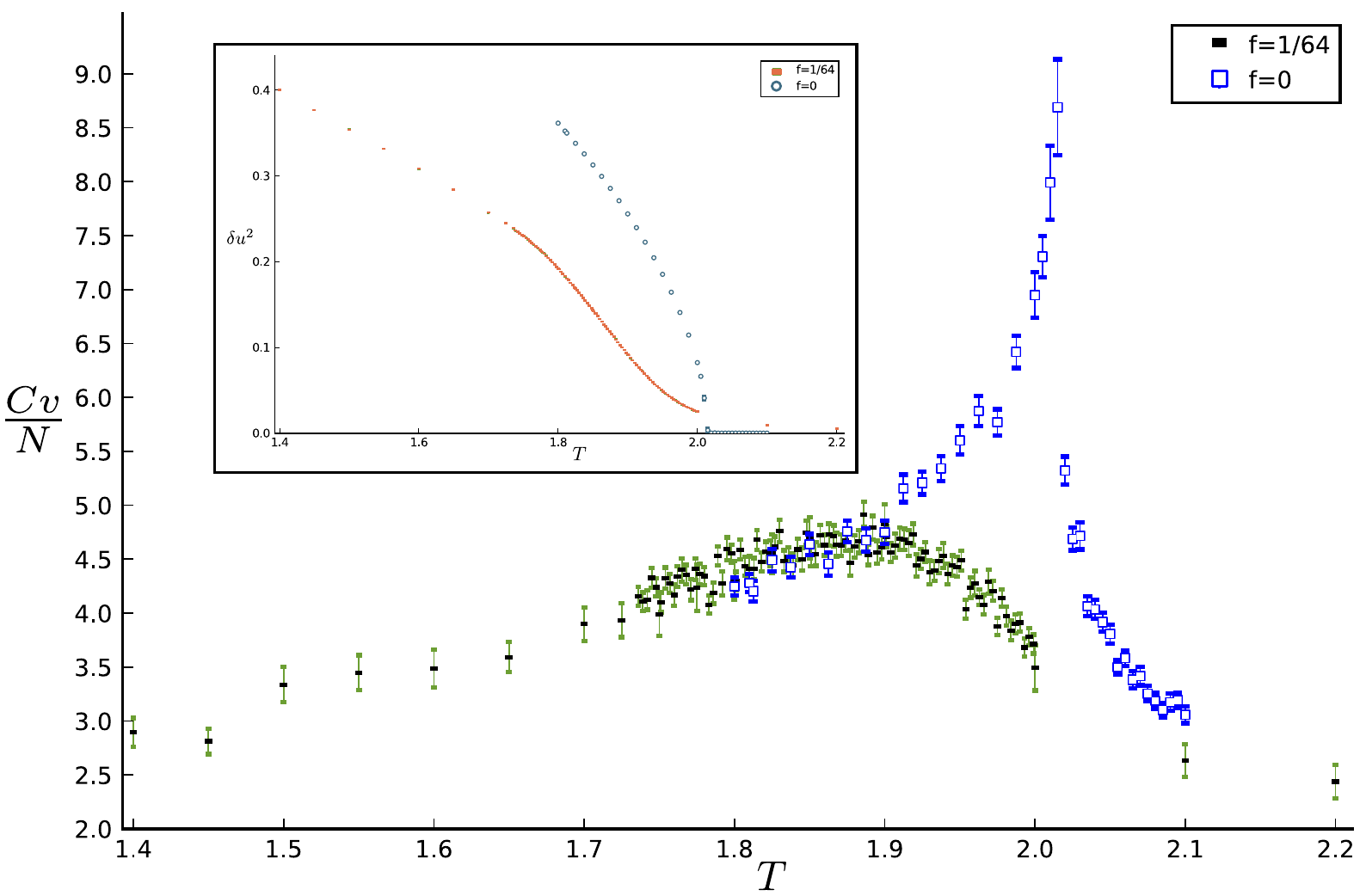}
    \caption{Specific heat dependence on temperature for a system with $g=0.3$, $\nu=0.1$ and $L=64$. 
        The blue points marked by a hollow square is for $f=0$, while the green data-set with points marked by a black dash is for a system with $f=1/64$. The inset shows the chiral order parameter $\delta u^2$ for the two filling fractions $f=0$ and $1/64$, with
        azure circles showing $f=0$
        while orange dashes show $f=1/64$. In the inset, the error-bars are for the most part too small to be seen.
    }
    \label{fig:specific_heat}
\end{figure}

To investigate what the relevant temperature-scale in our system is, we have performed Monte-Carlo simulations computing the specific heat and chiral order parameter at $f=0$ and $f=1/64$.
Fig.~\ref{fig:specific_heat} shows the specific heat as  a function of temperature at $f=0$. A sharp peak is seen at a temperature $T = 2.016\pm0.002$ and marks the phase transition from the superconducting to the normal state.
Also shown, is the specific heat at $f=1/64$, which at $T^* = 1.86\pm0.04$ shows a broadened and suppressed peak compared to $f=0$.
This peak marks the finite-field crossover to the normal state.
In what follows we will refer to this crossover temperature  as
 $T^*(f)$.
 
The inset shows the chiral order parameter as a function of $T$ at $f=0$ and $f=1/64$. 
For $f=0$, it vanishes at the same temperature as the sharp peak in the specific heat is located, and shows that the $f=0$ phase-transition in this model is associated with spontaneous time-reversal symmetry breaking. 
{For $f=1/64$, the presence of a magnetic field explicitly breaks time-reversal symmetry by selecting a preferred chirality, which leads to a finite order-parameter at $T^*(f)$.}

 These results form a useful background for choosing relevant temperatures at which to study vortex-lattice states at finite $f$. Below, we will study such vortex states in the temperature regime $T \in [1.5-1.8]$, and from the above results we conclude that these represent significant temperatures on the scale of the critical temperature $T_c$. Hence, our Monte-Carlo simulations at such temperatures will yield useful information concerning the thermal stability of the vortex states we find.       

\subsection{Vortex states upon lowering temperature}  \label{subsec:transition}

\subsubsection{Plasma State}\label{sec:PlasmaState}

\begin{figure}[h]
    \centering
    \includegraphics[width=\tableauWidth]{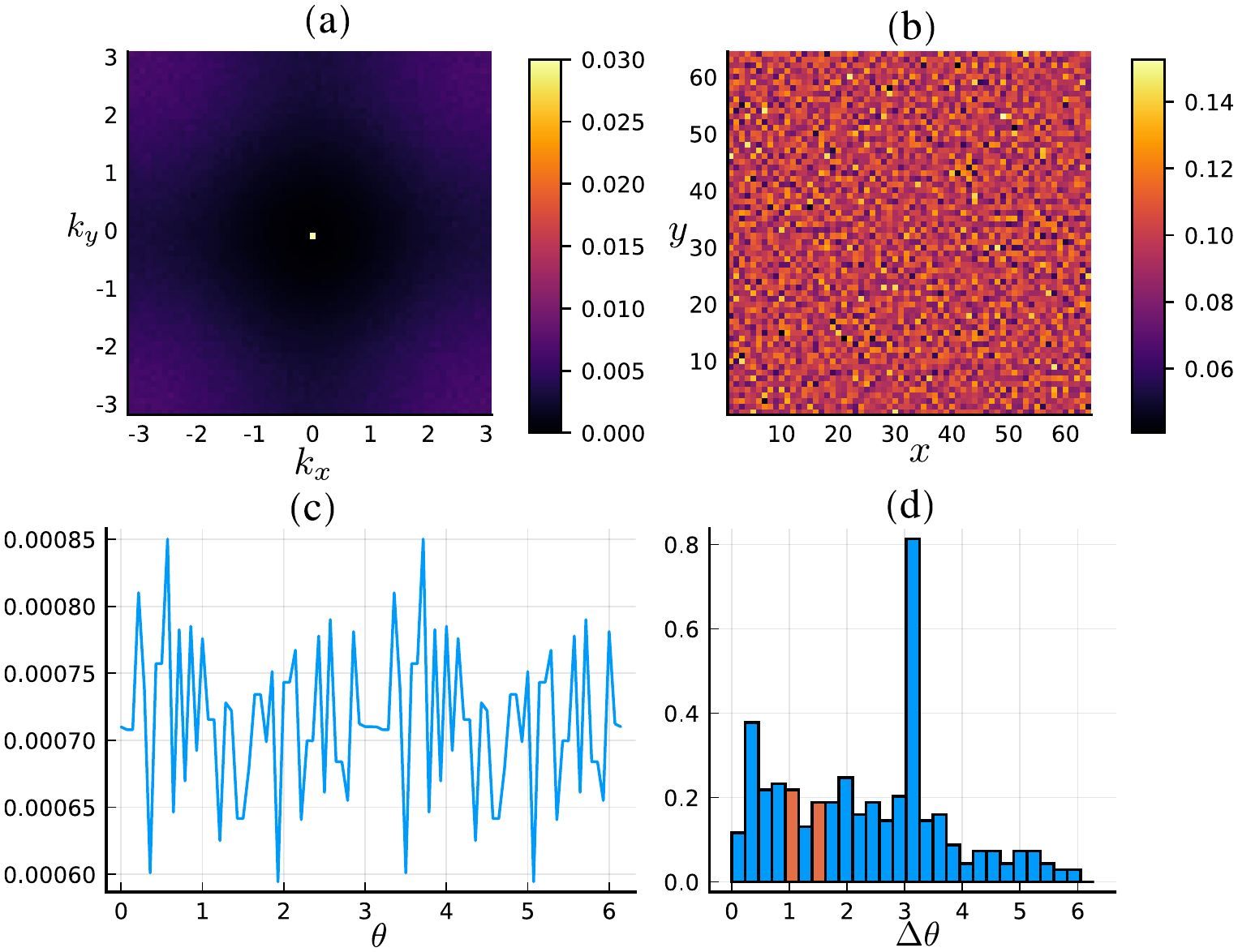}
    \caption{Vortex state at $T=2.0$ for a system with $\nu=0.1$, $g=0.3$ and $f = 1/64$. The system is dominated by thermally induced vortices. (a): Thermal average of the structure function. (b): Thermal average of real space vorticity. (c): Angular dependence of the structure function in a circular thin annulus around ${\bf k}=0$. (d): Histogram of angular difference $\Delta\theta$ between peaks in the angular dependence of the structure function. The colored bars are the bins that include $\Delta\theta = \pi/3$ and $\Delta\theta = \pi/2$. These would correspond to hexagonal and square lattices, respectively. } 
    \label{fig:PlasmaState:tableau}
\end{figure}

For $f=1/64$ and at high temperatures $T \agt 1.90$, the superconductor is in a normal state where thermal fluctuations have induced a proliferation of massive amounts of closed vortex-loops in the system.
The resulting state is therefore a vortex-plasma phase. This leads
to the tableau shown in Fig.~\ref{fig:PlasmaState:tableau}, which depicts results of simulations at $T=2.0$. The uniform distribution of vorticity in space leads to a circular
pattern at low $k$-vector magnitude with increasing value with increasing magnitude of the $k$-vector.
At higher $k$-vector magnitude, the value of the structure function exhibits a square anisotropy with
higher values close to $\v{k}_\text{corners}=\pi(1-2n, 1-2m)$ for $n,m\in\{0,1\}$. This anisotropy
is due to short range correlations since as $\v{k}$ approaches $\v{k}_\text{corners}$,
$\v{k}$ measures shorter and shorter correlations because of periodic boundary conditions. At these
length scales, the quadratic numerical lattice upon which the continuum model has been discretized gains
significance and leads to the apparent anisotropy.
The limits of the color bar reveal that this anisotropy is very small, with a maximum value
less than $0.010$.  There is no real signal of vortex-lattice correlations detected at this temperature.
The histogram in Fig.~\ref{fig:PlasmaState:tableau}d, reveals a large spike at $\Delta\theta = \pi$. This originates with the fact that the Fourier transform has the 
property $\mathcal{F}(\v{k}) = \mathcal{F}(-\v{k})^\ast$, such that the structure function is equal at $\v{k}$ and $-\v{k}$. 

\subsubsection{Singly-quantized square vortex lattice}
We next discuss the vortex lattice state that first emerges as the temperature is lowered below the crossover temperature to the normal state, which is $T^*=1.86$ at $\nu=0.1,g=0.3$, and $f=1/64$. 

\begin{figure}[h]
    \centering
    \includegraphics[width=\tableauWidth]{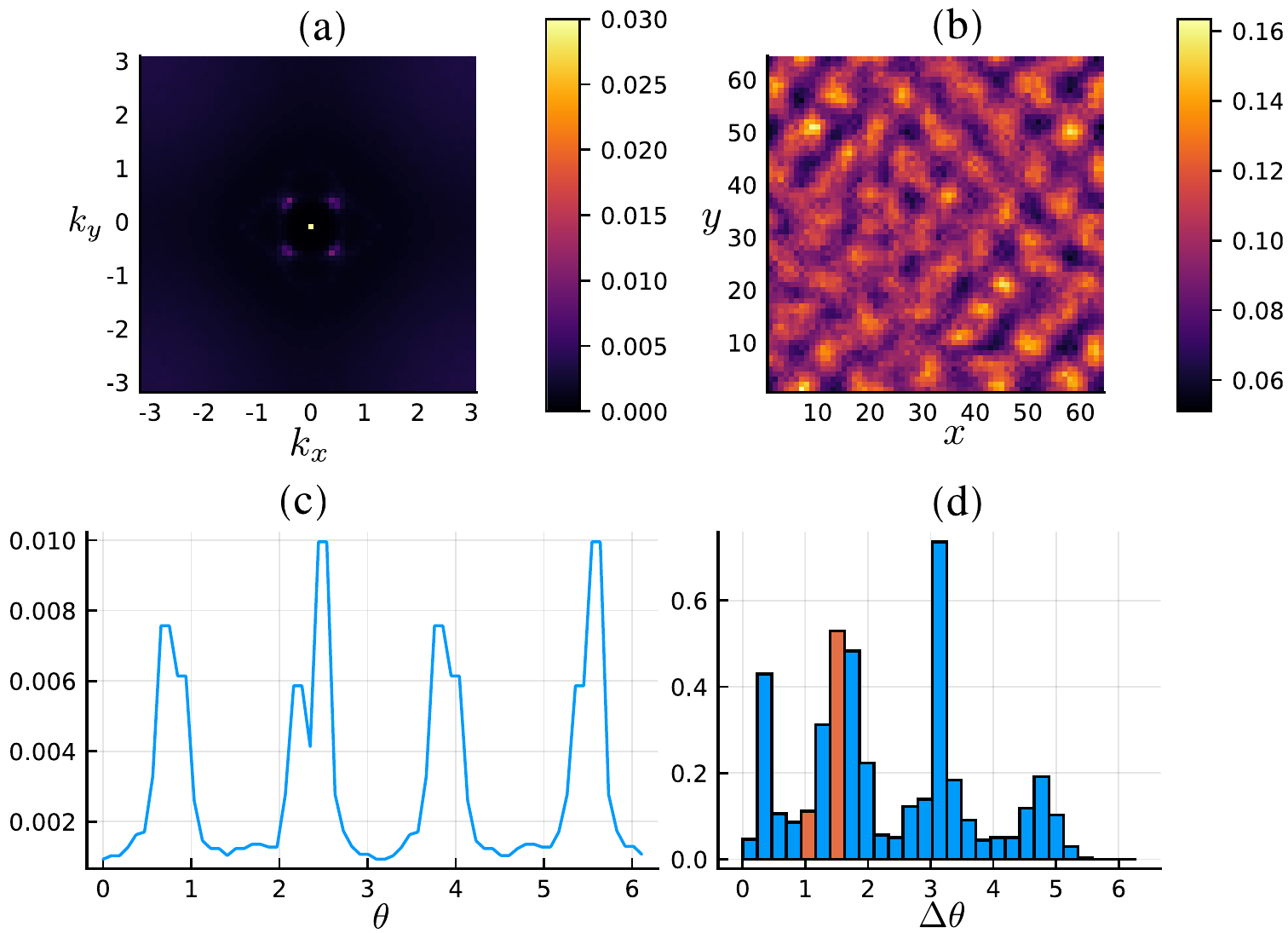}
    \caption{Singly-quantized square vortex lattice state for a system with $\nu=0.1$, $g=0.3$, $f = 1/64$, and $T=1.786$. (a)  shows the structure function of a square vortex lattice. (b) shows the vortex lattice structure in real-space. The vortices are located at the bright spots. All  vortices have a field-maximum at the center of the vortex, cf. Fig.~\ref{fig:schematic:1q:v}, consistent with singly-quantized vortices. 
    (c) shows the four-fold angular distribution of the structure function. (d) shows a histogram of the angular difference between peaks in the structure function. The colored histograms denote angular difference between peaks in the structure functions corresponding to $\pi/3$ and $\pi/2$. The dominant peaks are found at $\Delta \theta=  \pi/2,\pi$, and $3 \pi/2$, corresponding to a square lattice. 
    }
    \label{fig:tableau:sl}
\end{figure}

Fig.~\ref{fig:tableau:sl} shows the results of Monte-Carlo simulations performed at $T=1.786$, computing the structure function (a), vorticities (b), angular distribution of peaks in the structure function (c), and histograms of angular difference between peaks in the structure function (d). 
The structure function clearly has four-fold symmetry, such that the vortex lattice is square. This is also discernible in panel (b), although less obvious than in (a). The angular dependence of the structure function shown in (c)
shows four clear peaks separated by $\pi/2$. 
The histograms of $\Delta\theta$ in (d) shows that the most dominant non-trivial bin is $\pi/2$, marked by the orange bar. The broadening around the large orange bar is due to thermal fluctuations. The smaller orange bar represents the counts at angular difference of $\pi/3$, corresponding to a hexagonal lattice. The square lattice peak dominates the hexagonal peak, leading to the conclusion that the symmetry of the lattice is square, consistent with the result for the structure function in (a). The peak in (d) at low
angular value is attributed to the square lattice peaks being jagged due to the temperature being close to $T_c$.   

Fig.~\ref{fig:tableau:sl}b shows the square lattice structure as a real space average. One notable feature of the results of Fig.~\ref{fig:tableau:sl}, apart from the square vortex-lattice structure shown in (a), is that the magnetic field maximum associated with the vortices in (b) are located at the center of the vortices. Referring back to our discussion of Fig.~\ref{fig:schematic}, we see that this is consistent with singly-quantized vortices in each chiral component, $(n_+=1,n_-=-1)$.  

\begin{figure}[h]
    \subfloat[\label{fig:marked:vorticity}]{%
        \includegraphics[width=\markedPltWidth]{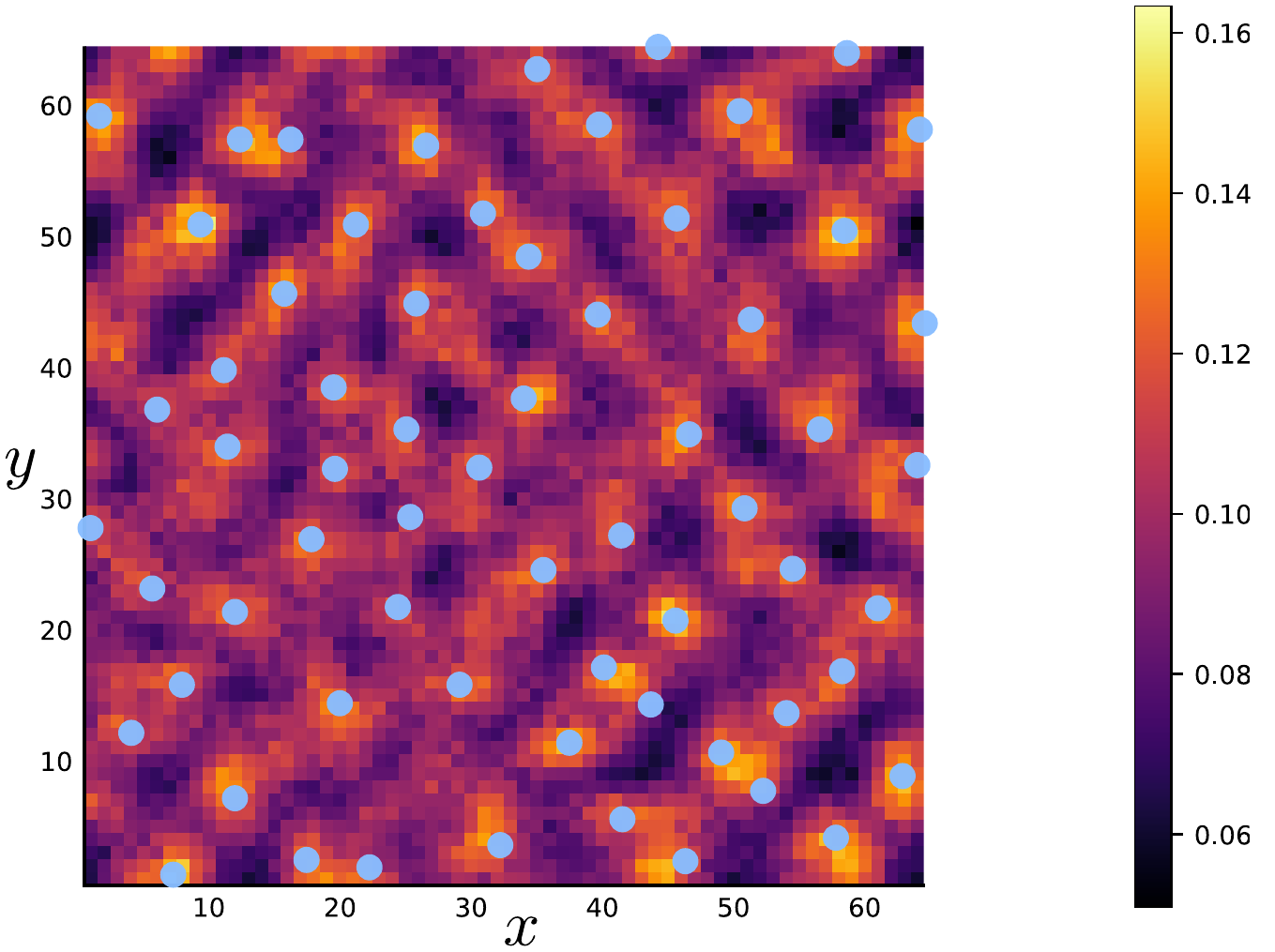}
    }
    \hfill
    \subfloat[\label{fig:marked:phaseDiff}]{%
        \includegraphics[width=\markedPltWidth]{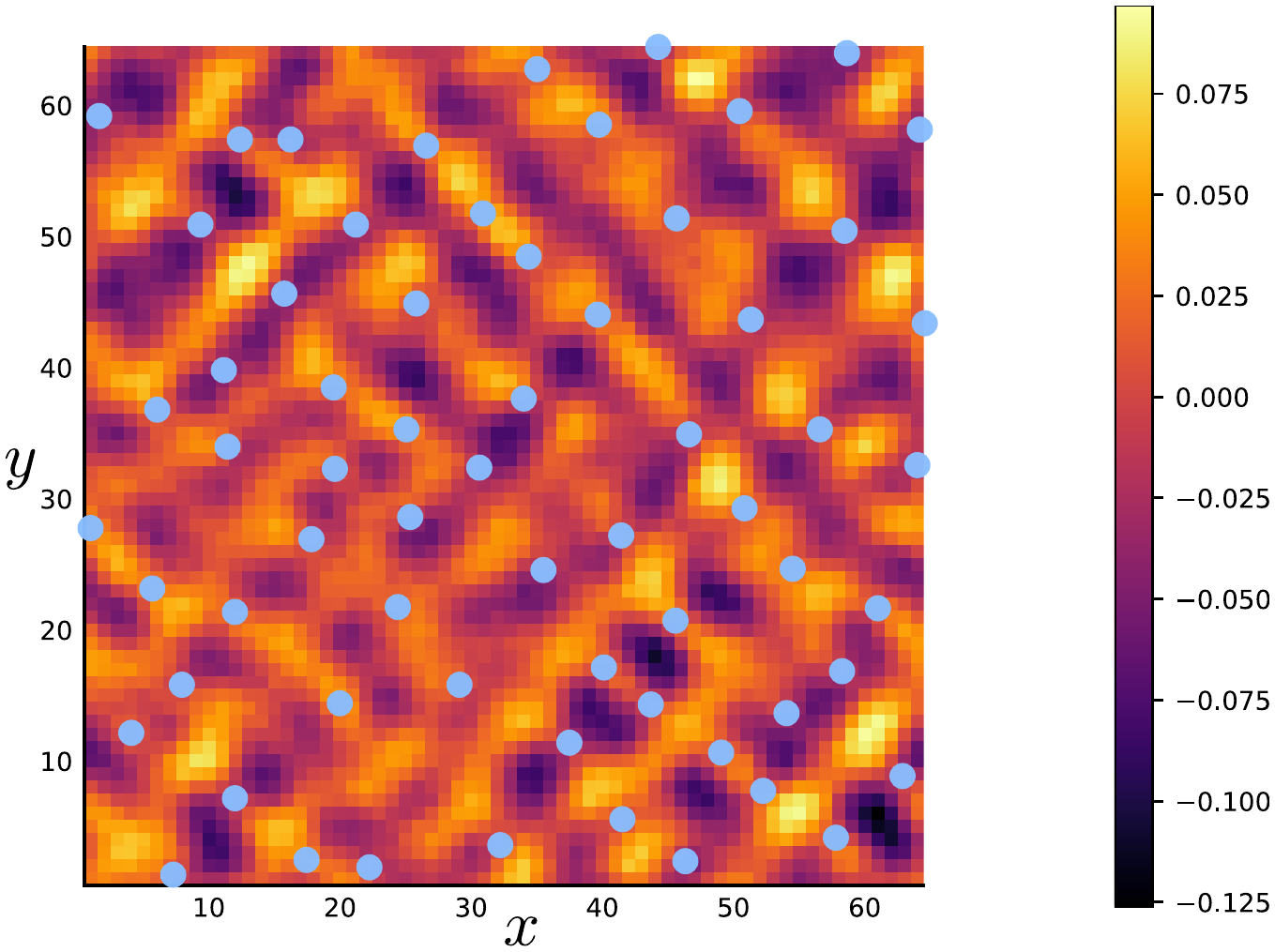}
    }
    \caption{Vortex positions and phase-differences for the parameters used in Fig.~\ref{fig:tableau:sl}. Fig.~\ref{fig:marked:vorticity} shows an enhanced version  of Fig.~\ref{fig:tableau:sl}~(b). The center of each vortex is marked by a green dot. The azure dots mark the positions of increased vorticity in the real space average. This corresponds to a maximum of the magnetic field  at the center of the vortex, cf. the schematics of Fig.~\ref{fig:schematic:1q:v}.
     Fig.~\ref{fig:marked:phaseDiff} shows the phase-difference around each vortex, whose position is indicated by a green dot. Note the four-fold symmetry of the phase-difference pattern around the vortices, cf. the schematics of 
     Fig.~\ref{fig:schematic:1q:pd}. Figs. \ref{fig:marked:vorticity} and 
     \ref{fig:marked:phaseDiff}
     corroborate, along with the results of Fig.~\ref{fig:tableau:sl} that at $(f=1/64,T=1.786)$, the vortex lattice is a singly-quantized square lattice.}
    \label{fig:marked}
\end{figure}

The nature of these points of increased vorticity is investigated further by comparing the position of these points with a real-space plot of average local phase-difference between the two components: $\langle\theta_\v{r}^+-\theta_\v{r}^-\rangle$, in Fig.~\ref{fig:marked}. The figures show that points of increased vorticity correspond well with intersections between two regions of positive average phase-difference and two
regions of negative average phase-difference. This corresponds to the the same phase-difference pattern that is depicted in Fig.~\ref{fig:schematic}c, again characteristic of singly-quantized vortices. 

The single quantum nature of the vortices is further corroborated by the fact that the boundary conditions enforce a total of $64$ quanta of magnetic flux at any step of the Monte-Carlo simulations. In Fig.~\ref{fig:tableau:sl}, there are $62$ clearly identifiable points of increased vorticity. It could be that the system shows $62$ single-quanta vortices and the remaining $2$ vortices are too thermally distorted to form enough of a coherent
thermal average to be identified,
or it could be that the system has $60$ single quanta vortices and $2$ double-quanta vortices. In any case, it is clear that the vortex state is dominated by singly-quantized vortices.  

\begin{figure}[h]
    \centering
    \subfloat[\label{fig:sl:amplitudes:u+}]{%
        \includegraphics[width=.23\textwidth]{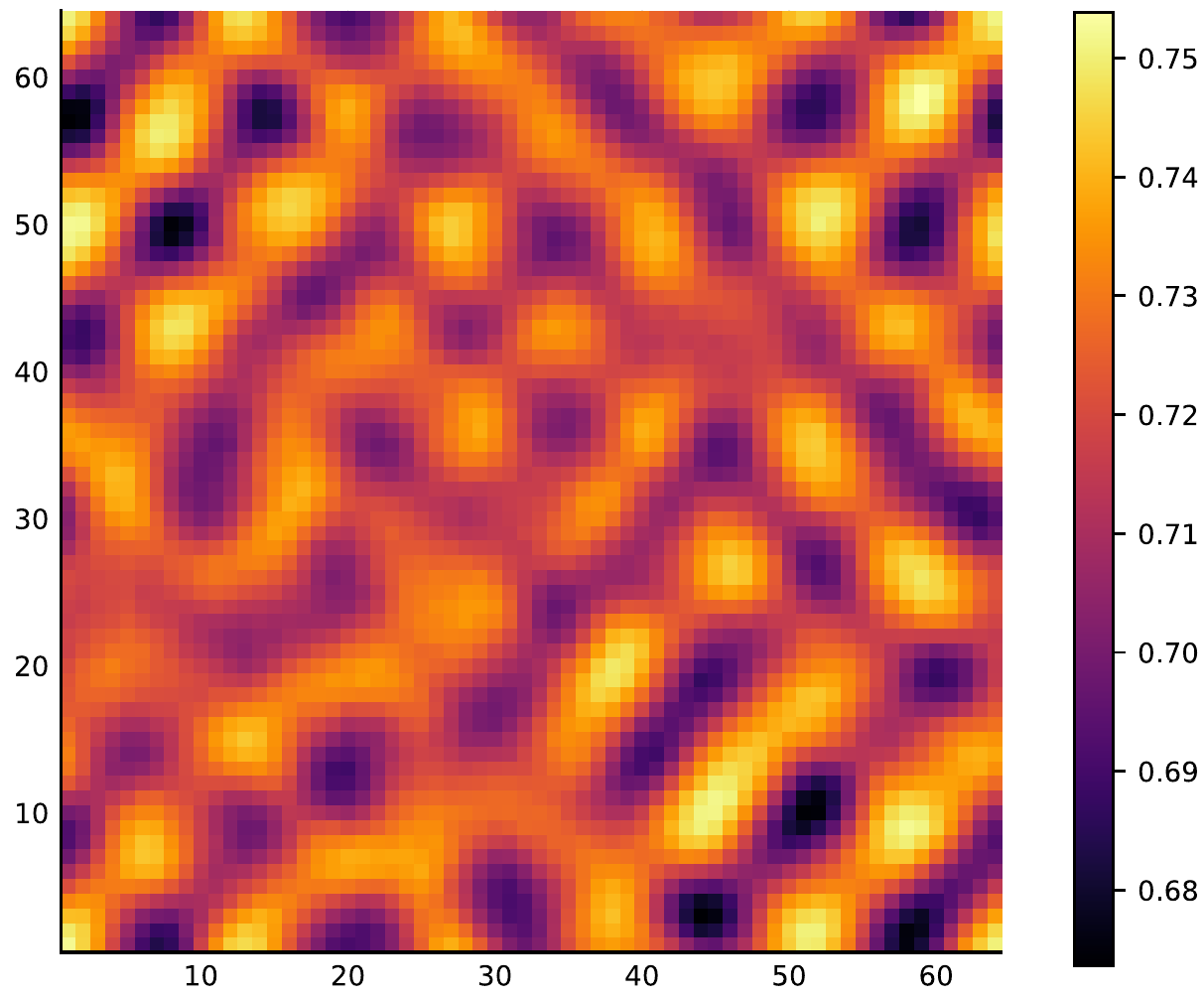}
    }\hfill
    \subfloat[\label{fig:sl:amplitudes:u-}]{%
        \includegraphics[width=.23\textwidth]{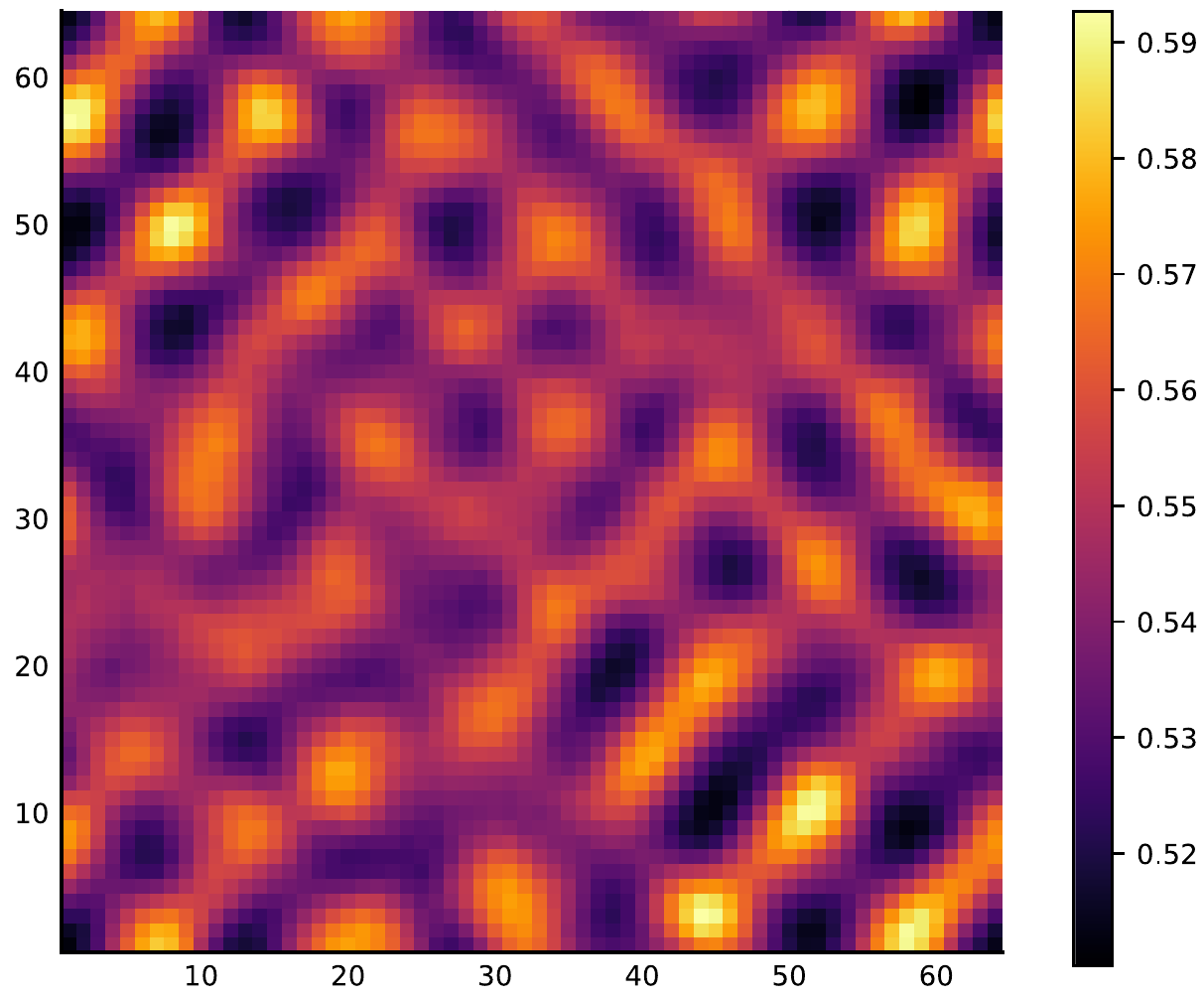}
    }
    \caption{Component amplitudes averaged in the $z$-direction for a system with $\nu=0.1$, $g=0.3$, $f = 1/64$ and $T=1.786$. \protect\subref{fig:sl:amplitudes:u+} shows $\langle\rho_{\v{r}_\perp}^+\rangle$ while \protect\subref{fig:sl:amplitudes:u-} shows $\langle\rho_{\v{r}_\perp}^-\rangle$. The color limits are set to amplify the spatial dependence, but we note that the average of $\rho^+$ is significantly higher than $\rho^-$.}
    \label{fig:sl:amplitudes}
\end{figure}

The superconducting field amplitude of conventional superconductors is suppressed in the presence of vortices. 
In the case of a two-component field, the sub-dominant component may be induced in the vicinity of the vortex core where the dominant component is suppressed \cite{AsleGaraud16}.
This is evident in Fig.~\ref{fig:sl:amplitudes} where the dominant component amplitude $\rho^+$ on the left exhibits dark regions that correspond
to the location of increased vorticity in Fig.~\ref{fig:tableau:sl}b and Fig.~\ref{fig:marked}a. 
On the right, the sub-dominant component exhibits increased amplitude in these regions as is required by the pseudo-$\mathbb{CP}^1$ constraint in Eq.~\eqref{eq:GLmodel:xy:CP1}.

We conclude from this that the stable vortex state at $\nu=0.1,g=0.3,f=1/64,T=1.786$ is a singly-quantized square vortex lattice. 

\subsubsection{Doubly-quantized hexagonal vortex lattice}
\begin{figure}[h]
    \centering
    \includegraphics[width=\tableauWidth]{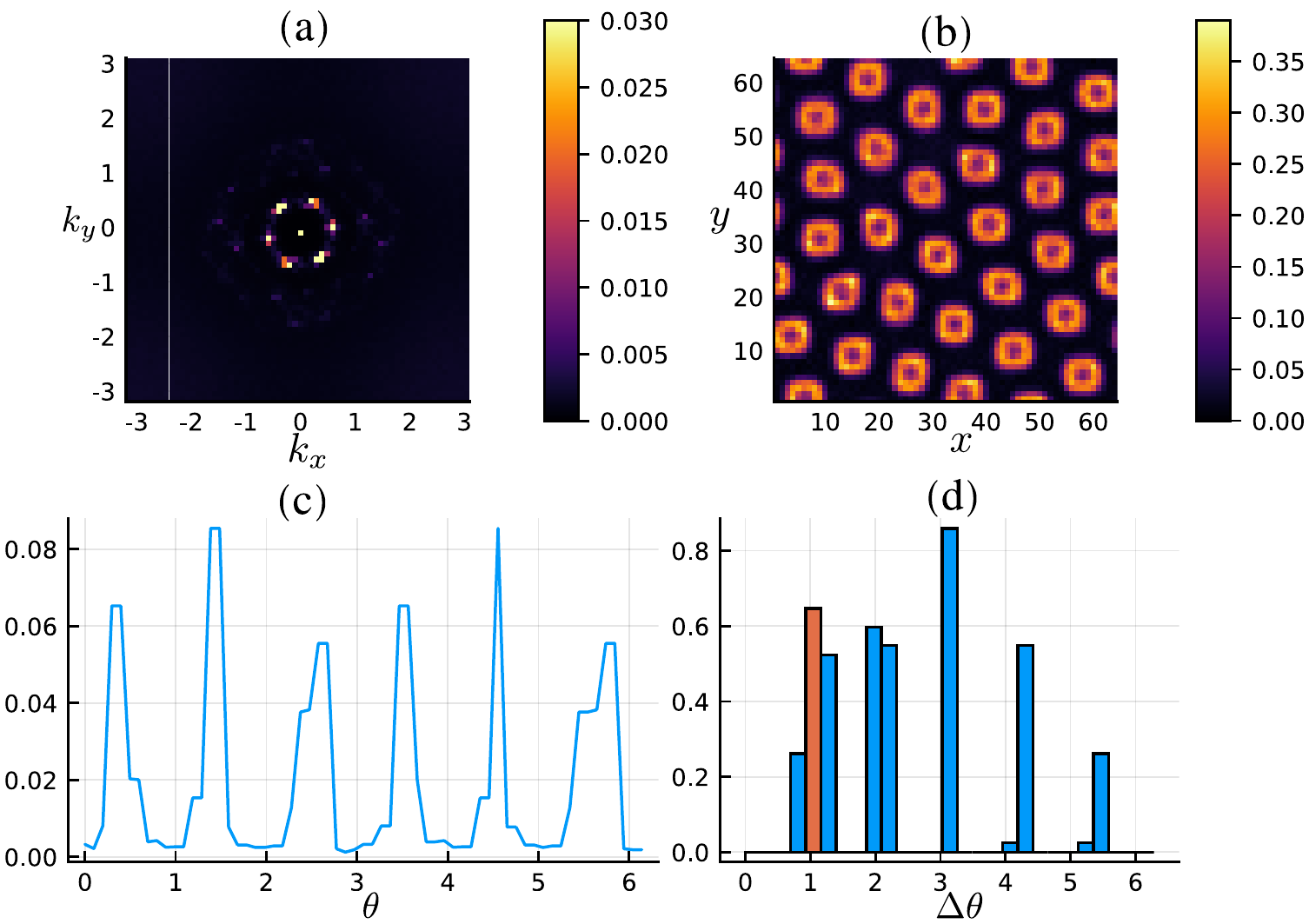}
    \caption{Doubly-quantized hexagonal vortex lattice state for a system with $\nu=0.1$, $g=0.3$, $f = 1/64$, and $T=1.5$. (a) shows the structure function, {showing a hexagonal lattice.} (b) shows the lattice structure in real-space. Vortices are located at the dark spots surrounded by a bright ring. All  vortices have a vorticity-maximum distributed in a ring around the center of the vortex, and a careful count shows that there are $32$ such doubly-quantized vortices, consistent with the system size $L_x \times L_y=64 \times 64$ and $f=1/64$. 
    This vortex-distribution is to be compared with the schematics of the upper right panel of Fig.~\ref{fig:schematic}.
    (c) shows the six-fold angular distribution of the structure function. (d) shows a histogram of the angular difference between peaks in the structure function. The colored histogram corresponds to an angular difference between peaks in the structure function of $\pi/3$. We see that the dominant peaks are found at $\Delta \theta=\pi/3$ and  $2\pi/3$, which corresponds to a hexagonal lattice. 
    }
    \label{fig:tableau:hl}
\end{figure}

We next consider the system at $f=1/64$ and a lower temperature $T=1.5$.
The plot of the average structure function  in Fig.~\ref{fig:tableau:hl}a shows $6$ clear, equidistantly placed peaks. Fig.~\ref{fig:tableau:hl}b, shows the average vorticities in real-space. The vorticity distribution around each vortex is clearly of the same type as depicted in  Fig.~\ref{fig:schematic:2q:v}, characteristic of doubly-quantized vortices. The angular dependence of the structure function in a thin annulus around ${\bf k} = 0$ is shown  
Fig.~\ref{fig:tableau:hl}c, where $6$ clear equidistantly placed peaks are seen. This is again reflected
in the  histogram for $\Delta \theta$ 
in Fig.~\ref{fig:tableau:hl}d
where a large peak is observed at $\Delta\theta = \pi/3$ followed by peaks at integer multiples of this. The real space vorticity average shows $32$ independent ring-structures (note that periodic boundary conditions have been used), which
indicates that each structure has $2$ quanta of magnetic flux.

\begin{figure}[h]
    \subfloat[\label{fig:HexLatt:marked:vorticity}]{%
        \includegraphics[width=\markedPltWidth]{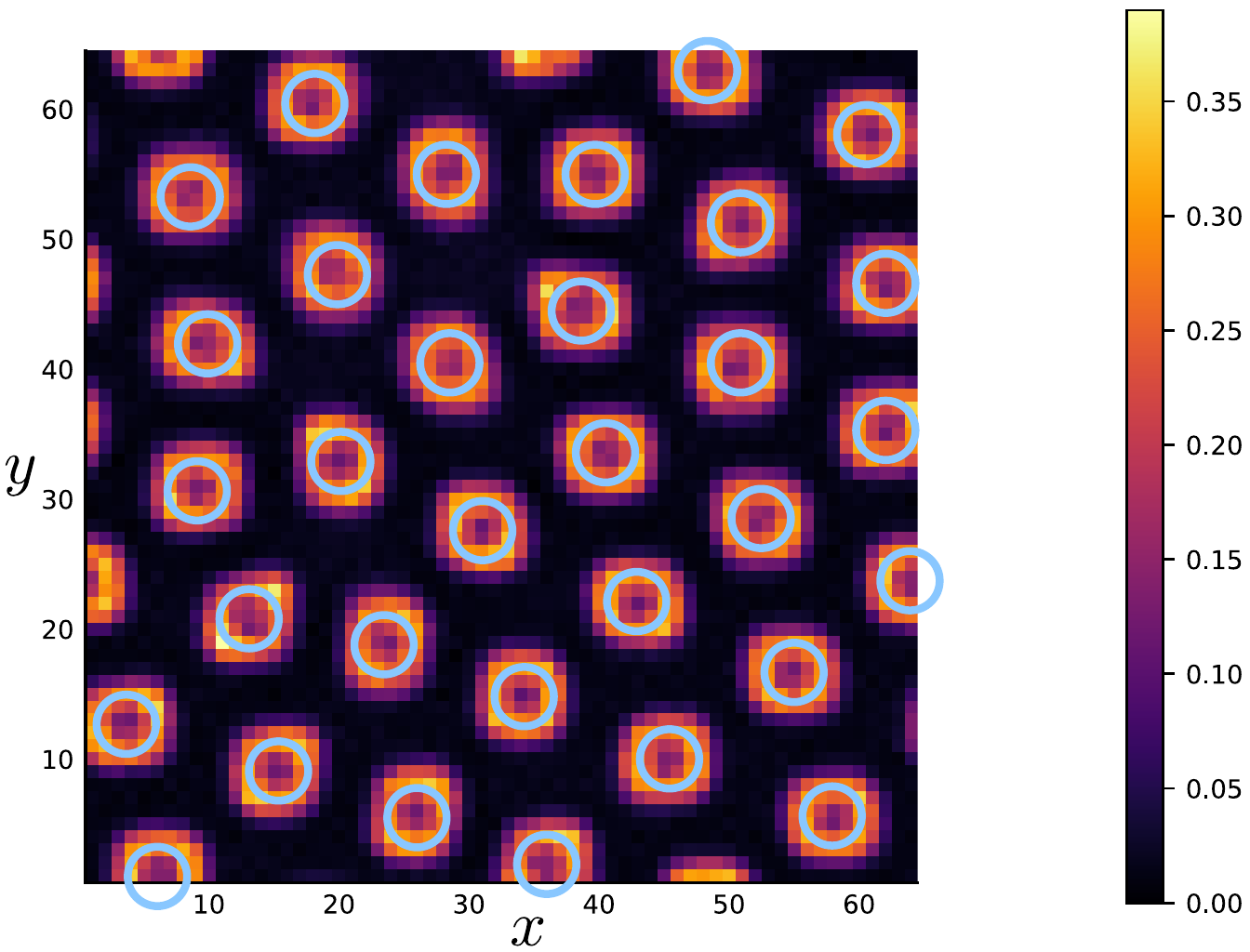}
    }
    \hfill
    \subfloat[\label{fig:HexLatt:marked:phaseDiff}]{%
        \includegraphics[width=\markedPltWidth]{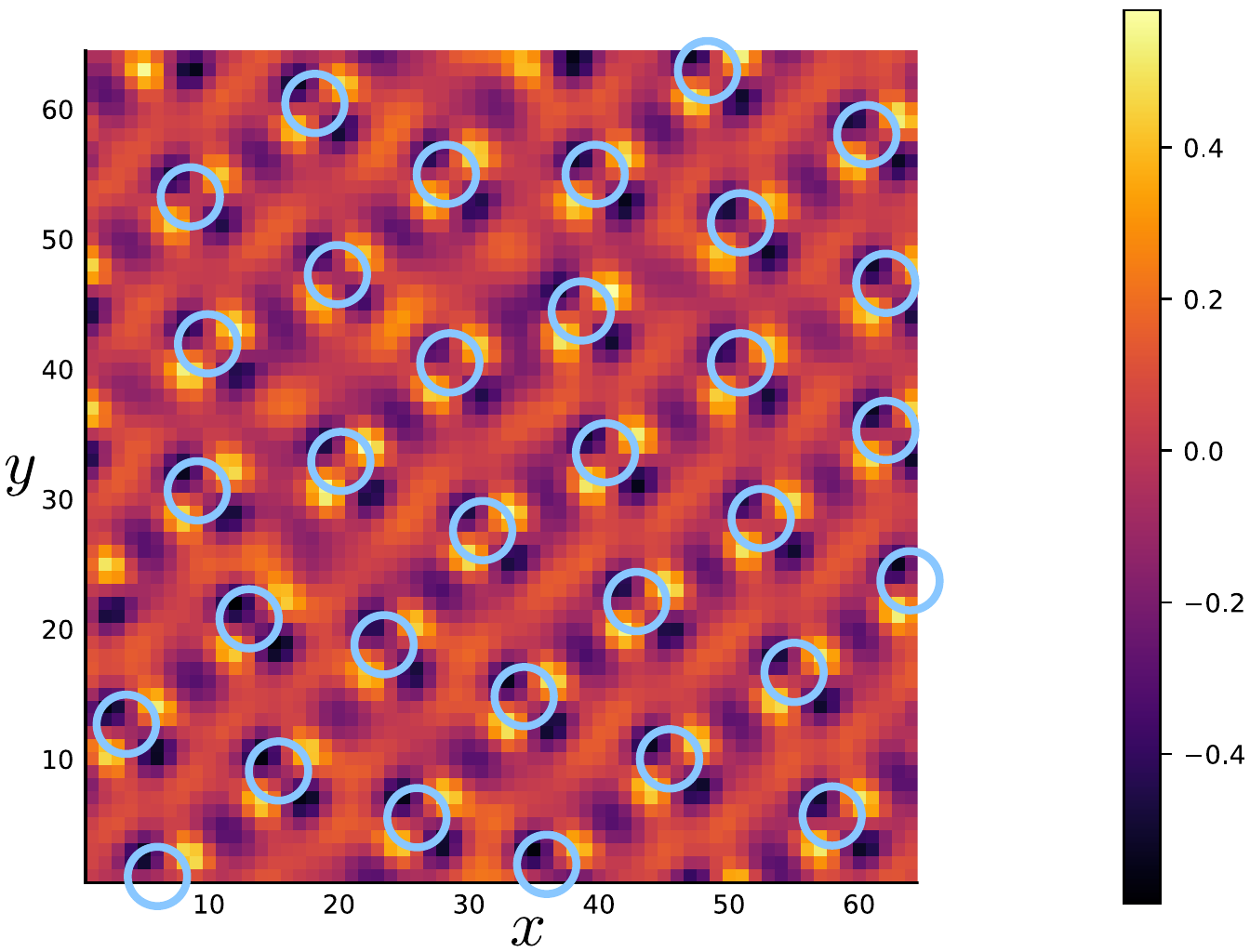}
    }
    \caption{Phase difference and $+$ vorticity of the system in Fig.~\ref{fig:tableau:hl}. The blue circles represent rings with increased vorticity in Fig.~\ref{fig:HexLatt:marked:vorticity}. These rings are then overlaid on the real space average of phase difference in Fig.~\ref{fig:HexLatt:marked:phaseDiff}.}
    \label{fig:HexLatt:marked}
\end{figure}

Fig.~\ref{fig:HexLatt:marked:vorticity} shows an enhanced version of Fig.~\ref{fig:tableau:hl}b. The ring-like structure of enhanced vorticity surrounding the center of each vortex is clearly seen, consistent with what is depicted in Fig.~\ref{fig:schematic:2q:v}. This is indicative of doubly-quantized vortices $(n_+=2,n_-=0)$. 
The double quantum nature of the vortices is also observed in the plot of real space phase difference average in Fig.~\ref{fig:HexLatt:marked:phaseDiff}.
It shows a clear inner $4\pi$ phase change at low radius from the vortex centre, where positive phase-difference is observed at an angle $\pi/4$
and $5\pi/4$ from the vortex centre and negative phase-difference at $3\pi/4$ and $7\pi/4$. 
This pattern is repeated at larger  radii away from the vortex core,  but then  rotated by $\pi/2$ degrees giving the vortices a distinct core structure not observed in the single-quantum case. It is finally noted that the real space average vorticity in Fig.~\ref{fig:tableau:hl} shows decreased vorticity in the vortex core for the positive component.

\begin{figure}[h]
    \centering
    \subfloat[\label{fig:tl:amplitudes:u+}]{%
        \includegraphics[width=.23\textwidth]{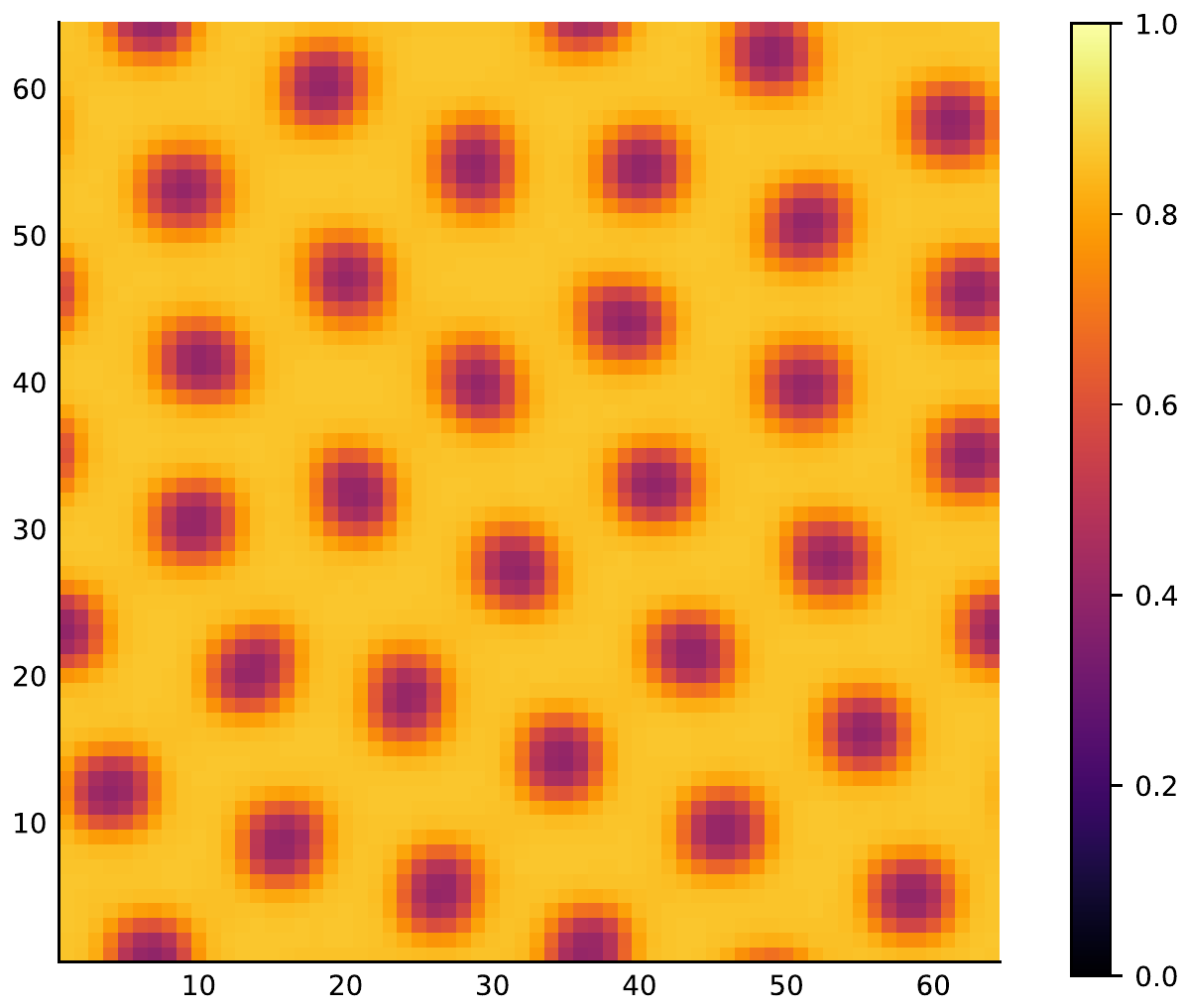}
    }\hfill
    \subfloat[\label{fig:tl:amplitudes:u-}]{%
        \includegraphics[width=.23\textwidth]{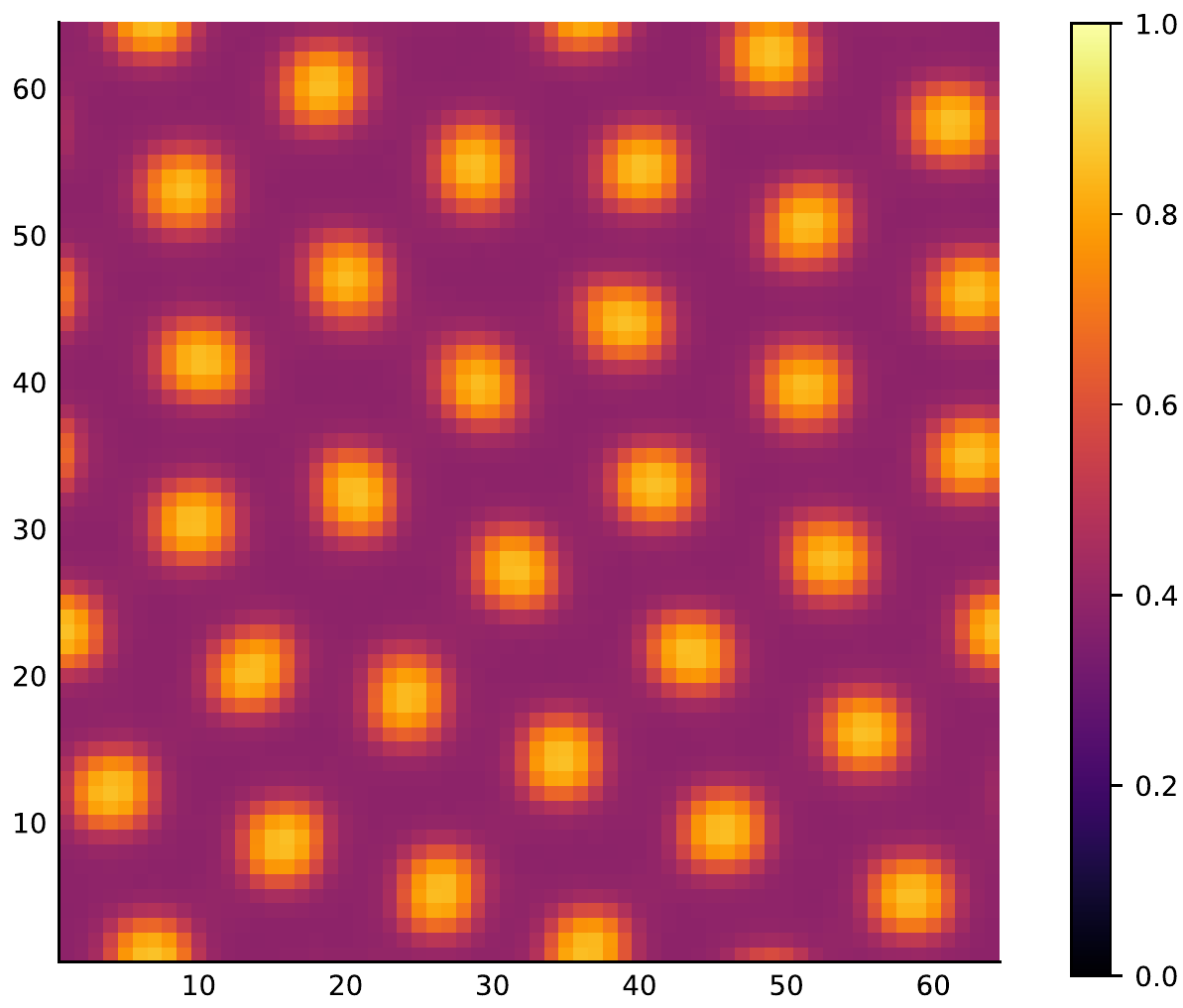}
    }
    \caption{Component amplitudes averaged in the $z$-direction for a system with $\nu=0.1$, $g=0.3$, $f = 1/64$ and $T=1.5$. \protect\subref{fig:sl:amplitudes:u+} shows $\langle\rho_{\v{r}_\perp}^+\rangle$ while \protect\subref{fig:sl:amplitudes:u-} shows $\langle\rho_{\v{r}_\perp}^-\rangle$. 
    In contrast to Fig.~\ref{fig:sl:amplitudes} the color limits are the same in both sub-plots since the lower temperature signal does not require amplification to discern spatial variance.}
    \label{fig:tl:amplitudes}
\end{figure}

The component amplitudes in Fig.~\ref{fig:tl:amplitudes} again reflect the hexagonal lattice pattern in Figs.~\ref{fig:tableau:hl} and \ref{fig:HexLatt:marked}. 
The dominant component on the left is clearly seen to be suppressed in the vicinity of the vortex cores, while the amplitude plot of the sub-dominant component on the right shows that this component in coincidently induced.

The conclusion is thus that the simulations at $f=1/64, T=1.5$ clearly show a hexagonal lattice of doubly-quantized vortices. 
Our simulations show that these doubly-quantized vortex states remain stable down to the lowest temperatures we have considered, and persist up to temperatures of $T=1.7$. The temperature-regime $T \in [1.7-1.75]$ will be discussed further below. 

\subsubsection{Mixed doubly and singly quantized vortex lattices}
\label{sec:mixedState}
\begin{figure*}[h]
    \centering
    \includegraphics[width=1.8\tableauWidth]{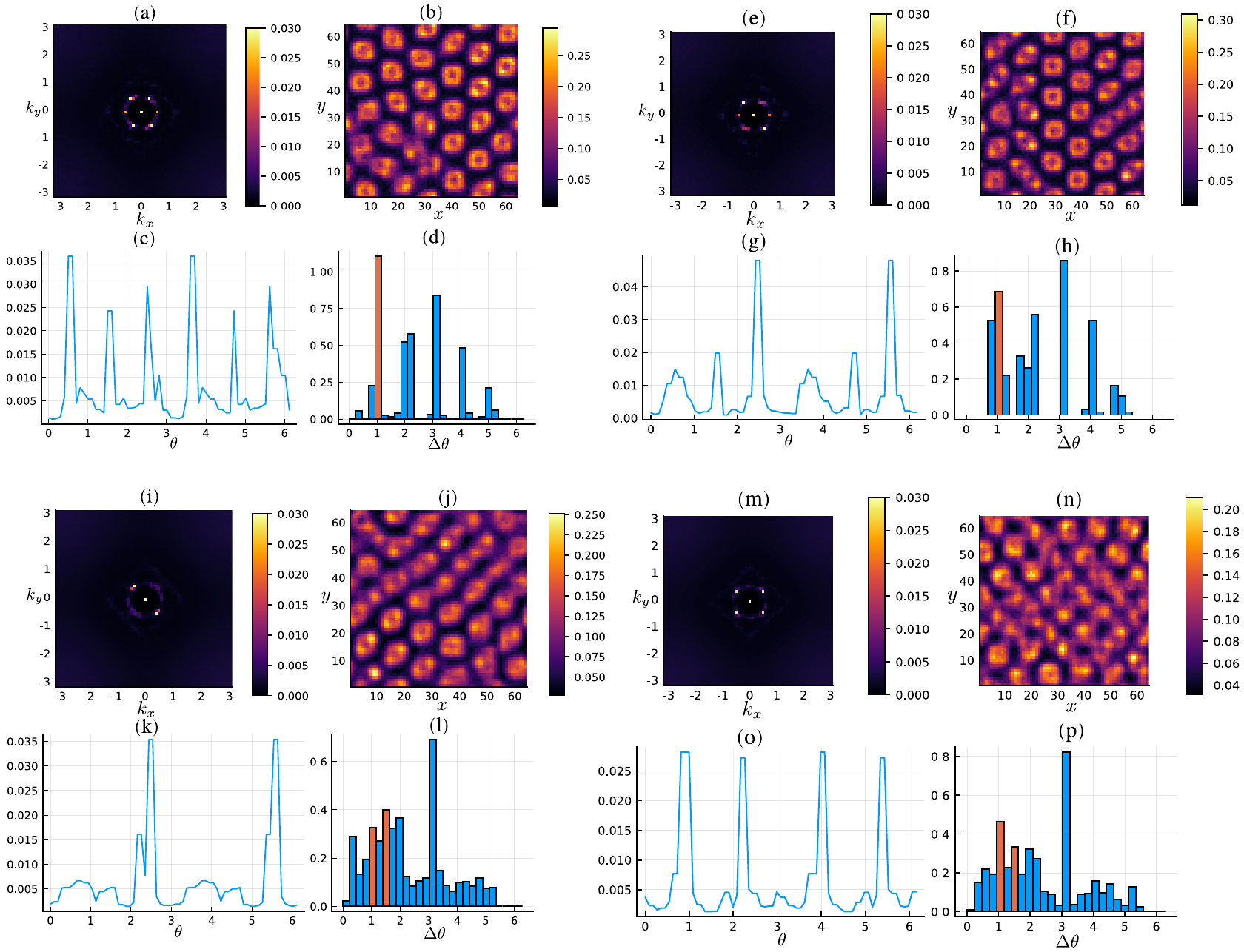}
    \caption{A tableau of simulation results for the temperatures $T=\{1.7, 1.725, 1.742, 1.751\}$ 
    in ascending order from left to right and top to bottom. 
The system has parameters $\nu=0.1$, $g=0.3$ and $f=1/64$. 
The real space $z$-averaged vorticity in (b), (f), (j) and (n) exhibits both single and double quanta lattice structures.
The remaining figures show the transition from signals of an hexagonal lattice to a square lattice as the temperature increases.}
    \label{fig:MixedState:quadTemp}
\end{figure*}

We next discuss the temperature regime where the transition from a higher-temperature singly-quantized square vortex lattice
to a lower-temperature doubly-quantized hexagonal vortex lattice takes place. 
For $f=1/64$, the transition takes place in the narrow range $T \in [1.73-1.775]$. 
Recall that the zero-field transition takes place at $T_c=2.016$ and the crossover temperature to the normal state at $f=1/64$ is $T \approx 1.86$. 
The four tableaus in Fig.~\ref{fig:MixedState:quadTemp} show examples of states of the system at intermediate temperatures $T=1.7,1.725,1.742,T=1.751$. 

At $T=1.7$ and $T=1.725$, the dominant structure is a doubly-quantized hexagonal vortex lattice.
The structure function of the vortex lattice is predominantly hexagonal,
see Fig.~\ref{fig:MixedState:quadTemp}~(a) and (e), but note the weakening of four of the peaks in the structure function in Fig.~\ref{fig:MixedState:quadTemp}~(e) compared to (a).
$\langle n^h_{\v{r}_\perp,z}\rangle$ in Figs.~\ref{fig:MixedState:quadTemp}~(b) and (f)
shows vortices characterized by a center with low vorticity surrounded by a ring of higher vorticity.
In this background, vortex structures start to appear that have a centre of high
vorticity, characteristic of singly-quantized vortices. 
Increasing  the temperature further, the hexagonal pattern in the structure function is gradually replaced by a square  pattern.

At $T=1.742$, the structure function features two strong peaks at opposite wave-vectors, with two weaker peaks in the orthogonal directions.
The overall symmetry of the structure function is now closer to one characteristic of a square lattice, see Fig.~\ref{fig:MixedState:quadTemp}i.
Namely, the four weaker spots in the six-fold symmetric structure functions in Fig.~\ref{fig:MixedState:quadTemp}~(a) and (e) have moved closer to each other. 
Although there is still a considerable number of doubly-quantized vortices present, 
i.e. vortices with low vorticity at the center surrounded by a ring of higher vorticity, 
it is evident that a substantial number of singly-quantized vortices have appeared, see Fig.~\ref{fig:MixedState:quadTemp}j.  

Increasing the temperature slightly to $T=1.751$, this becomes more pronounced. 
In Fig.~\ref{fig:MixedState:quadTemp}m, the four-fold symmetry of the structure function is evident, while  Fig.~\ref{fig:MixedState:quadTemp}n shows that there are still doubly-quantized vortices present. 
The transition from hexagonal to square vortex lattices upon increasing the temperature from $T=1.7$ to $T=1.75$, 
is mirrored in the peak-distance histogram with the bin at $\Delta\theta = \pi/3$ losing value and eventually being superseded by the bin at 
$\Delta\theta = \pi/2$, see Figs.~\ref{fig:MixedState:quadTemp}~(c), (g), (k) and (o), as well as Figs.~\ref{fig:MixedState:quadTemp}~(d), (h), (l) and (p).

\begin{figure}[h]
    \centering
    \includegraphics[width=0.5\textwidth]{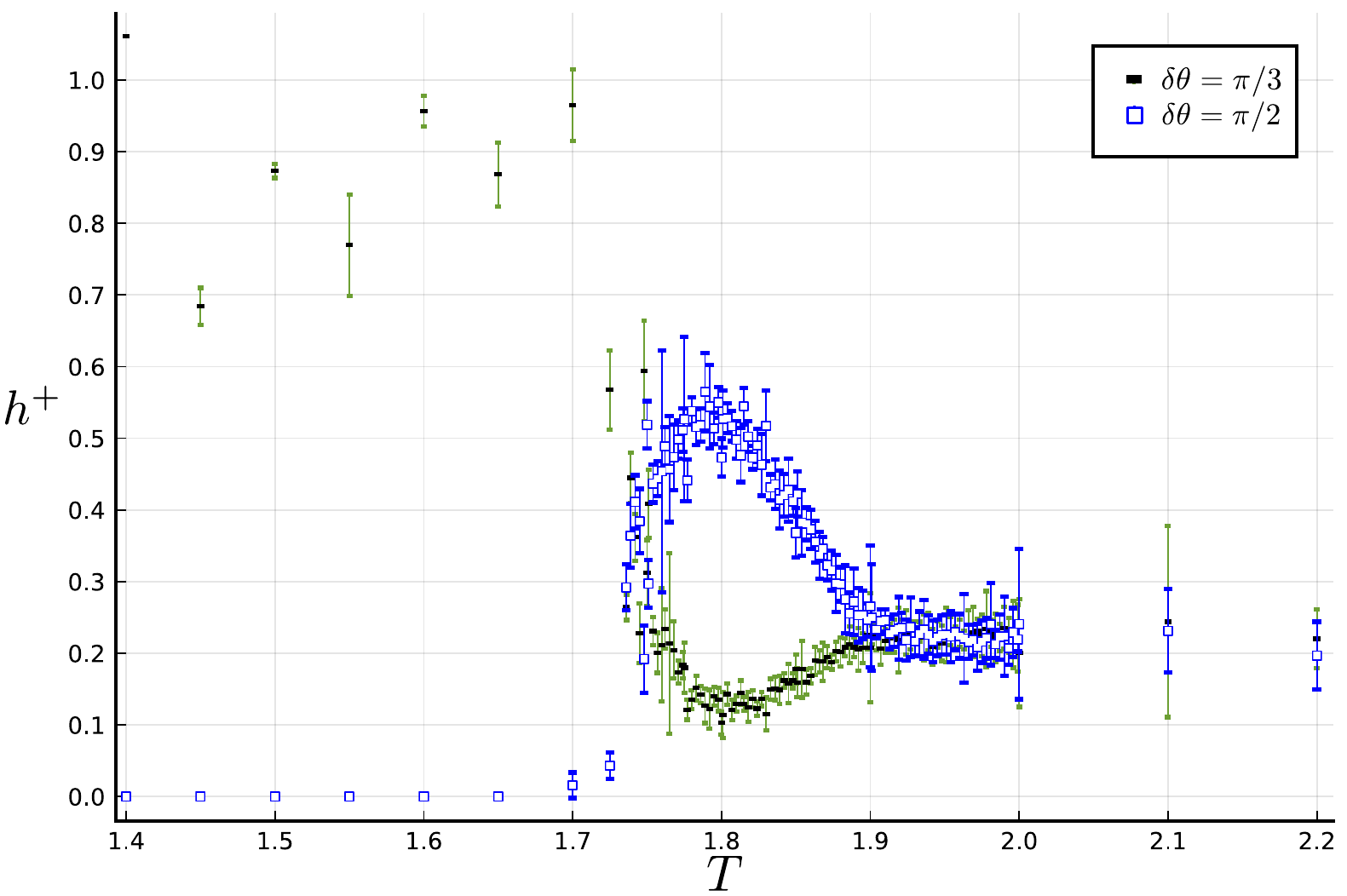}
    \caption{Histogram temperature dependence for two different bins in a system with $f=1/64$, $g=0.3$, $\nu=0.1$ and $L=64$. 
        The histogram is given by Eq.~\eqref{eq:GLmodel:Obs:hist} and gives normalized bins of the angular distance $\Delta\theta$ between peaks in the structure function.
        The bin at $\Delta\theta=\pi/2$ corresponds to the signal of a square lattice structure in the structure function and is marked with blue hollow squares. 
        The bin at $\Delta\theta=\pi/3$ is the signal for at hexagonal vortex lattice and is marked with black bar-markers and green error bars. The transition from a doubly-quantized hexagonal lattice to a singly-quantized square lattice as $T$ increases, occurs at $T \approx 1.75$.  
    }
    \label{fig:MixedState:histPlot}
\end{figure}

For a clearer picture of the temperature range over which this transition happens, we have computed the temperature dependence of these two histogram bins, shown in Fig.~\ref{fig:MixedState:histPlot}. The bin 
at $\Delta\theta=\pi/3$ (hexagonal vortex lattice) clearly dominates at lower temperatures, and becomes equal in height to the bin at $\Delta\theta=\pi/2$ (square vortex lattice) at $T \approx 1.75$. The temperature dependence of the two bins mirrors the dissociation of double quanta vortices into single quanta vortices which
we have noted is already starting at $T \approx 1.7$.
The histogram bins approach the value $h^+(\delta\Delta\theta) = 1/2\pi$ after the $U(1)$ crossover transition, which is an equal weight of the histogram on all bins. The lack of  angular variations in the structure function means that the system is in the vortex plasma phase.

The temperature regime in which the lattice reconstruction takes place is thus rather narrow and close to the $H_{c2}(T)$ crossover line. This is consistent with previous computations that ignored thermal fluctuations \cite{AsleGaraud16}, where the transition was induced by increasing the strength of the magnetic field up to values close to $H_{c2}$.

\section{Comparison with mean-field theory}
 Fig.~\ref{fig:MixedState:histPlot} gives a precise indication of where the vortex-lattice melting temperature in this system is, which is the temperature at which the two bins approach equal values and above  which their values remain constant. This occurs at $T^* \approx 1.88$, only slightly above the estimated $T$ where the broad peak in the specific heat is located at $T= 1.86$. 
The temperature window for which a square singly-quantized vortex lattice dominates is therefore conservatively estimated to be in the range $T \in [1.75-1.86]$. Below $T=1.7$, a doubly-quantized hexagonal vortex lattice is stabilized. Using the melting temperature as a measure of the transition to the normal state, i.e. as a measure of the upper critical field line, we see that the hexagonal lattice of double-quanta vortices is stable up to a temperature of about $0.9 T^*$.

\begin{figure}[h]
\centering
\includegraphics[width=0.5\textwidth]{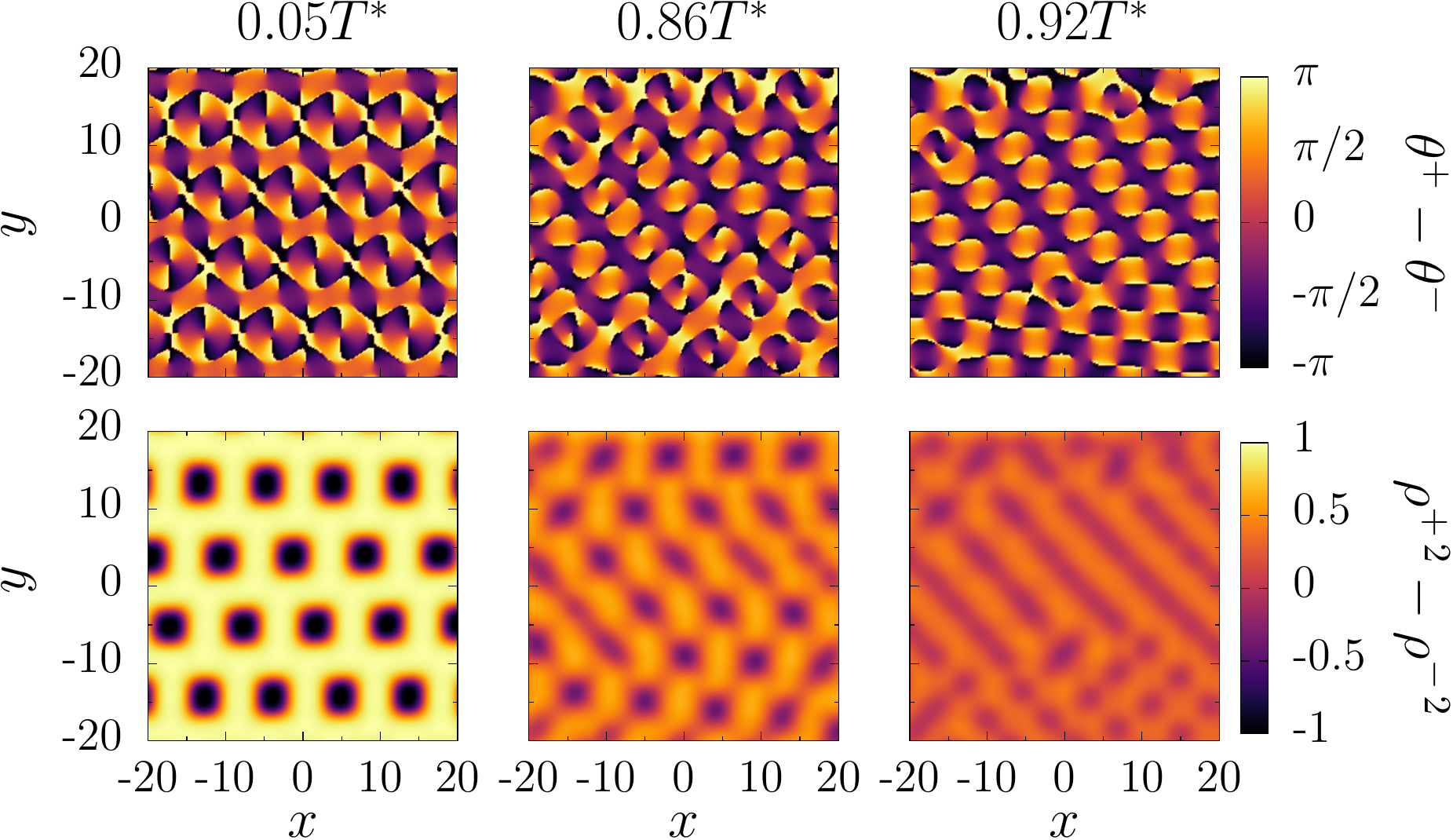}
\caption{Vortex phases, of the mean-field theory in an external field, for $g=0.3$ and 
$\nu=0.3$, with the corresponding filling fraction is $f_{MF}\approx1/30$.
The panels on the top line display the relative phase Eq.~\eqref{eq:GLmodel:Obs:phaseDiffAvg}
while the bottom line shows the relative densities defined in Eq.~\eqref{eq:GLmodel:Obs:du2}. 
Note that both relative phase and densities in the mean field are not thermal average.
}
\label{fig:MeanField}
\end{figure}

We now  compare these results quantitatively  with previously found mean-field results, where entropic effects were not fully accounted for \cite{AsleGaraud16}.
Fig.~\ref{fig:MeanField} displays three qualitatively 
different vortex phases obtained from simulations of mean-field theory, in an 
external field, at various temperatures. 
The procedure is to discretize the physical degrees of freedom $\eta_\pm$, and $\v{A}$
using a finite-element framework, and to numerically minimize the free energy 
\eqref{eq:GLmodel:chiral_clean_limit:potential}, in an external magnetic field 
(for details, see \cite{AsleGaraud16}).

The mean-field temperature is accounted for by modifying the quadratic 
term of the potential \eqref{eq:GLmodel:chiral_clean_limit:potential} to be 
$(T_{MF}-1)|\eta_h|^2$. There, the zero-field critical temperature is $T_{c,MF}=1$, 
and the crossover-line to the normal state at $f_{MF}\approx1/30$, is estimated, {from our numerical results} to be 
$T_{MF}^*=0.9$. 
For a better comparison of the role of the temperatures for the fluctuating theory 
with that of the mean-field, the results of the temperatures for mean-field simulations 
are expressed in unit of the crossover temperature $T_{MF}^*$.

At low temperatures, the minimal energy state is clearly an hexagonal lattice of 
double-quanta vortices. When approaching to the crossover temperature, the double-quanta 
vortices start to split into single-quanta vortices. Around $0.86T_{MF}^*$, there are few 
single quantum vortices, and the double-quanta still dominate. Closer to the crossover, 
around $0.92T_{MF}^*$, most of the vortices have dissociated and the single quantum 
vortices dominate. Eventually, the entire hexagonal lattice of double-quanta vortices 
has dissociated into a structure of single-quanta vortices. 
A generous estimate gives that the range of the mean-field temperature where the 
single-quanta vortices dominate to be about $0.1T_{MF}^*$. 
In general the entropic effects promote stability of the lattice
of single-quanta vortices. However, our Monte-Carlo simulations demonstrate that in the regime of parameters we have considered, the double-quanta vortex lattice is robust in a regime of temperatures approximately equal to what was found in previous work \cite{AsleGaraud16}.

%
%
%
%

\section{Summary}\label{sec:Summary}

In this paper, we have considered effects of thermal fluctuations on the vortex states in a model of a chiral $p$-wave superconductor with two complex matter-fields $(\eta^+,\eta^-)$ with opposite chiralities, for a filling fraction of $f=1/64$ vortices per square plaquette in the $(x,y)$-plane of a cubic numerical lattice, with an applied magnetic field in the $z$-direction. We have considered temperatures in the interval $T \in 1.5-2.0$, with the zero-field critical temperature ($f=0$) estimated to be $T_c=2.016\pm0.002$ and the crossover-line to the normal state at $f=1/64$, estimated to be $T^*=1.86\pm0.04$. 

At $T=1.5$ we have found that the stable field-induced vortex-configuration is a hexagonal vortex lattice of doubly-quantized vortices. At the higher temperature $T \approx 1.75$, this vortex lattice transitions, over a narrow temperature regime, to a square vortex lattice of singly-quantized vortices. At even higher temperatures, the vortex lattice structure function is washed out by thermally induced vortex loops when temperatures approach and cross the crossover-line at $f=1/64, T^* = 1.86\pm0.04$, rendering the system in a vortex-plasma phase. Our results indicate that double-quanta vortices can be  quite robust, and  do not very easily dissociate into single quanta vortices when thermal fluctuations are  included. 

Thus, previous results, based on ground state computations and minimization of internal energy, predicting  doubly-quantized hexagonal vortex lattices at low magnetic fields transitioning to singly-quantized square vortex lattices at higher magnetic fields very closely to $H_{c2}$, are stable to fully accounting for entropic effects in the free energy. 
 Therefore, double-quanta vortices is a quite robust property of chiral $p$-wave superconductors. 
 Our results, however, do indicate a slight broadening of the temperature regime above which a square vortex lattice is entropically stabilized compared to earlier mean-field results. The main finding is that, for the regimes considered in the paper, this entropic stabilization does not significantly diminish the temperature range where a doubly-quantized hexagonal vortex state exists. 
 \acknowledgements

F.N.K and A. S. would like to thank T.A. Bojesen for numerous invaluable discussions.
This work was supported by an NTNU University Grant,  the Research Council of Norway Project No. 250985 "Fundamentals of Low-dissipative Topological Matter", and the Research Council of Norway through its Centres of Excellence funding scheme, Project No. 262633, "QuSpin". E.B. was supported by the Swedish Research Council Grants No. 642-2013-7837, 2016-06122, 2018-03659, and G\"{o}ran Gustafsson Foundation for Research in Natural Sciences and Medicine and Olle Engkvists Stiftelse. We acknowledge support from the Norwegian High-Performance Computing Consortium NOTUR, Project NN2819K. 
\appendix

\section{Derivation of Eq. \eqref{eq:GLmodel:xy:chiralPhasesConstraint}}
\label{app:chiralPhases}
Writing $\tan\theta^\pm$ in terms of the complex fields yields
\begin{equation}
    \tan\theta^\pm = -i\frac{\eta^\pm - \eta^{\pm\,\ast}}{\eta^\pm + \eta^{\pm\,\ast}}.
    \label{eq:app:chiralPhases:tanByChiralFields}
\end{equation}
Inserting the transformation to the $xy$-basis in Eq.~\eqref{eq:GLmodel:chiral_transformation} and
making the London-limit approximation $|\eta^x|=|\eta^y|$, Eq.~\eqref{eq:app:chiralPhases:tanByChiralFields}
becomes
\begin{equation}
    \tan\theta^h = \frac{\sin\theta^x + h\cos\theta^y}{\cos\theta^x - h\sin\theta^y}.
    \label{eq:app:chiralPhases:tanByXYPhases}
\end{equation}
Using the trigonometric identity
\begin{equation}
    \sin x + \cos y = 2\sin\Big(\frac{x-y}{2} + \frac{\pi}{4}\Big)\sin\Big(\frac{x+y}{2} + \frac{\pi}{4}\Big),
    \label{eq:app:chiralPhases:trigID}
\end{equation}
after including $h$ and $-h$ in the argument of $\sin$ in the numerator and denominator of Eq.~\eqref{eq:app:chiralPhases:tanByXYPhases} respectively, $\tan\theta^h$ can be written
\begin{equation}
    \begin{split}
        \tan\theta^h = &h\frac{\sin\Big(\frac{h\theta^x-\theta^y}{2} + \frac{\pi}{4}\Big)\sin\Big(\frac{h\theta^x+\theta^y}{2} + \frac{\pi}{4}\Big)}{\sin\Big(-\frac{\theta^x+h\theta^y}{2}+\frac{\pi}{4}\Big)\sin\Big(\frac{\theta^x-h\theta^y}{2} + \frac{\pi}{4}\Big)}\\
        = &-h\frac{\sin\Big(\frac{\theta^x+\theta^y}{2}+h\frac{\pi}{4}\Big)}{\sin\Big(\frac{\theta^x+\theta^y}{2}-h\frac{\theta}{4}\Big)}\\
        = &-\frac{1}{-\bar{h}\frac{\sin\Big(\frac{\theta^x+\theta^y}{2} + \bar{h}\frac{\pi}{4}\Big)}{\sin\Big(\frac{\theta^x+\theta^y}{2} - \bar{h}\frac{\pi}{4}\Big)}} = -\frac{1}{\tan\theta^{\bar{h}}}.
    \end{split}
    \label{eq:app:chiralPhases:tanByPhaseSum}
\end{equation}
This equation shows that  both $\tan \theta^+$ and $\tan \theta^-$ are determined by one variable, $\theta^x+\theta^y$, which is what makes it possible to relate $\theta^+$ to $\theta^-$. 
Finally, by shifting the argument of the last $\tan$ we get Eq.~\eqref{eq:GLmodel:xy:chiralPhasesConstraint}, i.e. the relationship $\tan\theta^h = \tan(\theta^{\bar{h}} + \pi/2)$.

\begin{widetext}
\section{Symmetrized mixed gradient term}
\label{app:symm_MG}

Using the transformation properties of $\eta^a_{\v{r}}$ and $A_{\v{r},\mu}$ in 
Eq.~\eqref{eq:GLmodel:symm:Atransform} and \eqref{eq:GLmodel:symm:etaTrans} on the expression for discretized mixed gradient term in the $xy$-basis in
Eq.~\eqref{eq:GLmodel:xy:mgt} repeated here for convenience:
\begin{equation}
    \begin{split}
        \mathcal{F}^\mathrm{r}_\text{MG} = (1-\nu)\sum_a\Big[&\rho^a_{\v{r}+\hat{x}}\rho^{\bar{a}}_{\v{r}+\hat{y}}
            \cos\big(\theta^a_{\v{r}+\hat{x}} - \theta^{\bar{a}}_{\v{r}+\hat{y}} - (A_{\v{r},x}-A_{\v{r},y})\big)
            -\rho^a_{\v{r}+\hat{x}}\rho^{\bar{a}}_\v{r}\cos\big(\theta^a_{\v{r}+\hat{x}} - \theta^{\bar{a}}_\v{r} - A_{\v{r},x}\big)\\
        -&\rho^a_{\v{r}+\hat{y}}\rho^{\bar{a}}_\v{r}\cos\big(\theta^a_{\v{r}+\hat{y}} - \theta^{\bar{a}}_\v{r} - A_{\v{r},y}\big) 
    +\rho^a_\v{r}\rho^{\bar{a}}_\v{r}\cos\big(\theta^a_\v{r}-\theta^{\bar{a}}_\v{r}\big)\Big],
    \end{split}
    \label{eq:app:symm_MG:mgt}
\end{equation}
we obtain the rotated mixed gradient terms
\begin{align}
    \begin{split}
        C_4\mathcal{F}^\mathrm{r}_\text{MG} = -(1-\nu)\sum_a\Big[&\rho^a_{\v{r}-\hat{x}}\rho^{\bar{a}}_{\v{r}+\hat{y}}
            \cos\big(\theta^a_{\v{r}-\hat{x}} - \theta^{\bar{a}}_{\v{r}+\hat{y}} + (A_{\v{r},y}+A_{\v{r}-\hat{x},x})\big)
            -\rho^a_{\v{r}-\hat{x}}\rho^{\bar{a}}_\v{r}\cos\big(\theta^a_{\v{r}-\hat{x}} - \theta^{\bar{a}}_\v{r} + A_{\v{r}-\hat{x},x}\big)\\
            -&\rho^a_\v{r}\rho^{\bar{a}}_{\v{r}+\hat{y}}\cos\big(\theta^a_\v{r}-\theta^{\bar{a}}_{\v{r}+\hat{y}} + A_{\v{r},y}\big)
        +\rho^a_\v{r}\rho^{\bar{a}}_\v{r}\cos\big(\theta^a_\v{r}-\theta^{\bar{a}}_\v{r}\big)\Big],
    \end{split}
    \label{eq:app:symm_MG:C4mgt}\\
    \begin{split}
        C_4^2\mathcal{F}^\mathrm{r}_\text{MG} = (1-\nu)\sum_a\Big[&\rho^a_{\v{r}-\hat{x}}\rho^{\bar{a}}_{\v{r}-\hat{y}}
            \cos\big(\theta^a_{\v{r}-\hat{x}}-\theta^{\bar{a}}_{\v{r}-\hat{y}} - (A_{\v{r}-\hat{y},y}-A_{\v{r}-\hat{x},x})\big)
            -\rho^a_{\v{r}-\hat{x}}\rho^{\bar{a}}_\v{r}\cos\big(\theta^a_{\v{r}-\hat{x}}-\theta^{\bar{a}}_\v{r}+A_{\v{r}-\hat{x},x}\big)\\
            -&\rho^a_\v{r}\rho^{\bar{a}}_{\v{r}-\hat{y}}\cos\big(\theta^{\bar{a}}_{\v{r}-\hat{y}}-\theta^a_\v{r}+A_{\v{r}-\hat{y},y}\big)
        +\rho^a_\v{r}\rho^{\bar{a}}_\v{r}\cos\big(\theta^a_\v{r}-\theta^{\bar{a}}_\v{r}\big)\Big],
    \end{split}
    \label{eq:app:symm_MG:C42mgt}\\
    \begin{split}
        C_4^3\mathcal{F}^\mathrm{r}_\text{MG} = -(1-\nu)\sum_a\Big[&\rho^{\bar{a}}_{\v{r}-\hat{y}}\rho^a_{\v{r}+\hat{x}}
            \cos\big(\theta^{\bar{a}}_{\v{r}-\hat{y}}-\theta^a_{\v{r}+\hat{x}} + (A_{\v{r}-\hat{y},y} + A_{\v{r},x})\big)
            -\rho^{\bar{a}}_{\v{r}-\hat{y}}\rho^a_\v{r}\cos\big(\theta^{\bar{a}}_{\v{r}-\hat{y}}-\theta^a_\v{r}+A_{\v{r}-\hat{y},y}\big)\\
            -&\rho^{\bar{a}}_\v{r}\rho^a_{\v{r}+\hat{x}}\cos\big(\theta^a_{\v{r}+\hat{x}}-\theta^{\bar{a}}_\v{r}-A_{\v{r},x}\big)
        +\rho^{\bar{a}}_\v{r}\rho^a_\v{r}\cos\big(\theta^{\bar{a}}_\v{r}-\theta^a_\v{r}\big)\Big].
    \end{split}
    \label{eq:app:symm_MG:C43mgt}
\end{align}
In these expressions, $a,q\in\{x,y\}$. Adding Eqs.~\eqref{eq:app:symm_MG:mgt} -  \eqref{eq:app:symm_MG:C43mgt},  several terms cancel. 
As is immediately obvious, all the on-site terms such as the last term in Eq.~\eqref{eq:app:symm_MG:mgt} cancel each other.
Considering the last term on the first line of Eq.~\eqref{eq:app:symm_MG:C4mgt}, we let $\v{r}\to\v{r}+\hat{x}$ which is allowed because of
periodic boundary conditions, and we see that this cancels the last term on the first line of Eq.~\eqref{eq:app:symm_MG:mgt}.
The first term on the last line of Eqs.~\eqref{eq:app:symm_MG:C4mgt} and \eqref{eq:app:symm_MG:mgt} can be seen to cancel through a simple re-labeling
of the $a$ summation index. 
The same cancellations happen for the analogous terms in Eqs.~\eqref{eq:app:symm_MG:C42mgt} and \eqref{eq:app:symm_MG:C43mgt} such that the average
of Eqs.~\eqref{eq:app:symm_MG:mgt}-\eqref{eq:app:symm_MG:C43mgt} and thus the full symmetrized expression for the mixed gradient terms 
can be written on the simple form
\begin{equation}
    \mathcal{F}^\mathrm{s}_\text{MG} = \frac{(1-\nu)}{4}\sum_a\sum_{h,h^\prime=\pm}hh^\prime\rho^a_{\v{r}+h\hat{x}}\rho^{\bar{a}}_{\v{r}+h^\prime\hat{y}}
    \cos\big(\theta^a_{\v{r}+h\hat{x}} - \theta^{\bar{a}}_{\v{r}+h^\prime\hat{y}} - A_{\v{r},hx} + A_{\v{r},h^\prime y}\big).
    \label{eq:app:symm_MG:symmMG}
\end{equation}
This expression, together with Eqs.~\eqref{eq:GLmodel:Disc:gauge}, \eqref{eq:GLmodel:xy:kinetic}, \eqref{eq:GLmodel:xy:potential} and
\eqref{eq:GLmodel:xy:anisotropy}
 together constitute the free energy used in the simulations.
\end{widetext}

\section{Numerical basis rotation}
\label{app:phaseConversion}
In this appendix, we present the numerical details for how chiral matter field-amplitudes and -phases are calculated from their $xy$-basis counterparts.

The chiral amplitudes $\rho^h_\v{r}$ are easily found from the $xy$-basis variables through Eq.~\eqref{eq:GLmodel:xy:chiralAmplitudes}:
\begin{equation}
    \rho^{h}_\v{r} = \sqrt{\frac{\rho^{x\,2} + \rho^{y\,2}}{2} +h \rho^x\rho^y\sin\big(\theta^x_\v{r}-\theta^y_\v{r})}.
    \label{eq:GLmodel:Obs:amp}
\end{equation}
The chiral phases
are obtained by the set of equations
\begin{align}
    \sin\theta^h_\v{r} &= \frac{\rho^x\sin\theta^x_\v{r}+h\rho^y\cos\theta^y_\v{r}}{\sqrt{2}\rho^h_\v{r}},
    \label{eq:GLmodel:Obs:phases:sin}\\
    \cos\theta^h_\v{r} &= \frac{\rho^x\cos\theta^x_\v{r}-h\rho^y\sin\theta^y_\v{r}}{\sqrt{2}\rho^h_\v{r}}.
    \label{eq:GLmodel:Obs:phases:cos}
\end{align}
As long as $\rho^h_\v{r} > 0$, $\theta^h_\v{r}\in[-\pi,\pi)$ can be found through simple trigonometric relations which we include for completeness.
Given that $\cos\theta^h_\v{r} > 0$ then $\theta^h_\v{r} = \tan^{-1}\tan\theta^h_\v{r}$.
If $\cos\theta^h_\v{r} < 0 $ then $\theta^h_\v{r} = \tan^{-1}\tan\theta^h_\v{r} - \pi\sgn\tan\theta^h_\v{r}$.
The final case is that $\cos\theta^h_\v{r} = 0$ in which case $\theta^h_\v{r} = \pi/2\sgn\sin\theta^h_\v{r}$.

In the chiral ground state of the system $\theta^x_\v{r}-\theta^y_\v{r}\to -h\pi/2$ which makes $\rho^h_\v{r}\to0$ when $\rho^x=\rho^y$. 
This makes Eqs.~\eqref{eq:GLmodel:Obs:phases:sin} and \eqref{eq:GLmodel:Obs:phases:cos} numerically unstable as both numerator and denominator approach zero.
To accurately calculate $\theta^h_\v{r}$, these equations are expanded around the ground state value.
Setting $\theta^x_\v{r}-\theta^y_\v{r} = -h\pi/2+2\pi n + \delta$, and expanding to $4$th order in $\delta$ yields
\begin{subequations}
    \begin{align}
        \begin{split}
            \sin\theta^h_\v{r} \to &\frac{\delta}{\abs{\delta}}\cos\theta^x_\v{r}\Big[1 - \frac{\delta^2}{8} + \frac{\delta^4}{384}\Big]\\
            &-\frac{\abs{\delta}}{2}\sin\theta^x_\v{r}\Big[1 - \frac{\delta^2}{24} + \frac{\delta^4}{1920}\Big],
        \end{split}
        \label{eq:GLmodel:Obs:expand:sin}\\
        \begin{split}
            \cos\theta^h_\v{r} \to 
            &\frac{\abs{\delta}}{2}\cos\theta^x_\v{r}\Big[1-\frac{\delta^2}{24} + \frac{\delta^4}{1920}\Big] \\
            &-\frac{\delta}{\abs{\delta}}\sin\theta^x_\v{r}\Big[1 - \frac{\delta^2}{8} + \frac{\delta^4}{384}\Big].
        \end{split}
        \label{eq:GLmodel:Obs:expand:cos}
    \end{align}
    \label{eq:GLmodel:Obs:expand}
\end{subequations}
The expressions on the right are independent of $h$.
Then if $\sin\theta^h_\v{r} \leq 0$, $\theta^h_\v{r} = -\cos^{-1}\cos\theta^h_\v{r}$. If not, then $\theta^h_\v{r} = \cos^{-1}\cos\theta^h_\v{r}$.
To find $\delta$ we simply calculate $\delta = \mod(\theta^x_\v{r}-\theta^y_\v{r},2\pi) - 3\pi/2$ for $h=+$ and
$\delta = \mod(\theta^x_\v{r}-\theta^y_\v{r},2\pi) -\pi/2$ for $h=-$.
With this expansion in $\delta$, the errors from calculating $\theta^h_\v{r}$ were found to be smaller than the floating point error.

\bibliography{probib}

\end{document}